%% file: Main_arxiv_2.tex
\documentclass[trackchanges]{aastex7}

\usepackage{graphicx} 

\definecolor{xkcdbrightblue}{RGB}{43, 103, 252}

\begin{document}
\title[Panchromatic Spectra of Nearby Stars]{Panchromatic Spectra of Nearby Low-mass and Sun-like Stars with Directly Imageable Habitable Zones}

\author[0000-0002-1046-025X]{Sarah Peacock} 
\affiliation{University of Maryland, Baltimore County, MD 21250, USA} 
\affiliation{NASA Goddard Space Flight Center, Greenbelt, MD 20771, USA}
\email{sarah.r.peacock@nasa.gov}

\author[0000-0002-4955-0471]{Breanna A. Binder}
\affiliation{Department of Physics and Astronomy, California State Polytechnic University, Pomona, CA, USA}
\email{babinder@cpp.edu}

\author[0000-0002-2949-2163]{Edward W. Schwieterman}
\affiliation{Department of Earth and Planetary Sciences, University of California, Riverside, CA, USA}
\affiliation{Blue Marble Space Institute of Science, Seattle, WA, USA}
\email{eschwiet@ucr.edu}

\author[0000-0002-0569-1643]{Margaret C. Turnbull}
\affiliation{SETI Institute, Carl Sagan Center for the Study of Life in the Universe, Off-Site: Madison, WI 53713, USA}
\email{turnbull.maggie@gmail.com}

\author[0000-0002-7084-0529]{Stephen R. Kane}
\affiliation{Department of Earth and Planetary Sciences, University of California, Riverside, CA, USA}
\email{skane@ucr.edu}

\author[0000-0001-6398-8755]{Katherine Garcia-Sage}
\affiliation{NASA Goddard Space Flight Center, Greenbelt, MD 20771, USA}
\email{katherine.garcia-sage@nasa.gov}

\author[0000-0003-0944-2334]{Alison Farrish}
\affiliation{NASA Goddard Space Flight Center, Greenbelt, MD 20771, USA}
\email{alison.o.farrish@nasa.gov}
\affiliation{George Mason University, Fairfax, VA, USA}

\begin{abstract}
Characterizing the high-energy radiation environments of stars plays a critical role in determining which systems' planets are capable of retaining atmospheres and sustaining habitable conditions. X-ray through ultraviolet (UV) radiation drives atmospheric photochemistry, heating, and escape, making accurate characterization of stellar high-energy emission essential for both interpreting future exoplanet observations and identifying the most promising targets for life detection. We construct panchromatic spectral energy distributions (SEDs) spanning the X-ray through radio for 12 nearby low-mass and Sun-like stars with directly imageable habitable zones that are prioritized targets for the Habitable Worlds Observatory (HWO) and Extremely Large Telescopes (ELTs). These SEDs are generated using forward stellar atmosphere models guided and constrained by available archival X-ray and UV observations. We find that many stars in this sample exhibit elevated high-energy radiation environments relative to the modern Sun, with habitable zone X-ray+extreme UV (XUV) fluxes frequently exceeding solar values by 1–2 orders of magnitude. The elevated emission likely reflects a combination of sample selection effects, differences in stellar age and rotation, and intrinsic magnetic variability, with multi-epoch observations demonstrating that variability alone can significantly alter inferred radiation environments. These results highlight high-energy radiation as an important discriminator in identifying the most promising habitable planet hosts and demonstrate the need for expanded X-ray and UV observations to complete the stellar characterization necessary for HWO target prioritization.
\end{abstract}


\section{Introduction}

We are entering a new era of exoplanet science in which the direct imaging and characterization of potentially habitable worlds is becoming a near-term reality, driven by the forthcoming Habitable Worlds Observatory (HWO) and the Extremely Large Telescopes (ELTs). Together, these facilities will enable reflected-light detections and spectroscopic measurements of Earth-sized planets orbiting nearby stars. The subset of stars with habitable zones (HZs) that are directly imageable, set by a combination of stellar brightness, distance, and angular separation \citep{Turnbull15,mamajek2023}, represents a distinct and strategically important population for both HWO and ELT programs. These systems are among the most valuable targets for assessing planetary atmospheres and surface conditions in the search for habitable environments beyond the Solar System. To interpret any future detections or spectra from these planets, it is essential to first understand their host stars in detail, as the stellar spectral energy distribution (SED) defines the radiation field that governs planetary climate, chemistry, and potential biosignatures.

Accurate knowledge of a star's SED is fundamental for interpreting exoplanet observations, as stellar radiation defines the planetary energy budget, climate state, and atmospheric chemistry. The stellar flux determines the balance between heating and cooling that sets a planet's equilibrium temperature and directly influences habitability through climate forcing \citep[e.g.,][]{Kasting1993, Shields2013, Kopparapu2014, Kane2016, Kopparapu2017,Eager2020,Hill2023}. It also drives atmospheric escape and photochemical processes that control the composition and stability of planetary atmospheres \citep[e.g.,][]{Tian2005, MurrayClay2009, Owen2012, Ginzburg2018, Kubyshkina2018, Howe2020, Johnstone2021}. In addition, variability and heterogeneity in stellar emission can imprint signatures that mimic or obscure planetary features, complicating the interpretation of reflected-light or transmission spectra \citep[e.g.,][]{Pont2013,McCullough2014,Oshagh2014,Llama2015,Rackham2018,Rackham2023}. Together, these factors mean that the properties of the host star (e.g., luminosity, spectral shape, and activity) govern nearly every aspect of exoplanet characterization, from establishing the planet's equilibrium temperature and assessing its potential habitability, to interpreting atmospheric features and evaluating the impact of stellar contamination on observed spectra.

Panchromatic spectra, spanning from X-rays through the infrared and into the radio, capture the complete radiative environment experienced by orbiting planets. Accurate models of photochemistry, atmospheric escape, and surface UV conditions on potentially habitable exoplanets depend sensitively on the spectral shape and intensity of stellar X-ray through near-ultraviolet (NUV) emission \citep[e.g.,][]{Harman2015,Meadows2018b, teal2022}. Each region of the stellar spectrum encodes distinct physical processes and plays a unique role in shaping planetary atmospheres. X-rays ($<$100\,\AA) trace high-energy coronal emission that heats and ionizes upper atmospheres, driving atmospheric loss \citep[e.g.,][]{lopez2012, Owen2012}. Extreme ultraviolet (EUV; 100--912\,\AA) photons provide the dominant energy input for atmospheric escape and ionospheric chemistry, though they are currently inaccessible with operational telescopes \citep[e.g.,][]{lammer2007,MurrayClay2009, Koskinen2010,Chadney2015,Johnstone2019}. Far-ultraviolet (FUV; 912--1800\,\AA) flux controls the photodissociation of key molecules such as H$_2$O, CO$_2$, H$_2$, and CH$_4$, and includes critical diagnostics of stellar magnetic activity \citep[e.g.,][]{trainer2006, Zerkle2012,Arney2017}. NUV (1800--3200\,\AA) photons regulate ozone chemistry and link high-energy variability to photospheric output \citep[e.g.,][]{Seager2000, segura2010, Hu2012, Harman2015, tilley2019}. The visible spectrum (0.32--1 $\mu$m) contains the bulk of the photospheric emission for Sun-like stars and is essential for modeling planetary albedo and reflected-light spectra \citep[e.g.,][]{Marley1999,Seager2005,Feng2018}. Infrared (IR; 1--1000 $\mu$m) flux constrains thermal emission and provides sensitivity to cool starspots and faculae \citep[e.g.,][]{Barnes2011,Iyer2020}. Finally, radio emission ($>$1000 $\mu$m) traces stellar magnetic fields, winds, and large-scale particle acceleration events \citep[e.g.,][]{Hallinan2015,Vedantham2020}.

Panchromatic SEDs have been assembled for the Sun and for dozens of nearby stars in past studies, advancing our understanding of the radiation environments that exoplanets experience. These include solar reference spectra and major multiwavelength surveys such as the MUSCLES, Mega-MUSCLES, and MEATS Treasury programs \citep{France2016, Loyd2016, Behr2023, Wilson2025a, Wilson2026}, as well as single-star analyses \citep[e.g.,][]{Fontenla2016, Peacock2019b, Hintz2019, Feinstein2022,Diamond-Lowe2024}. Collectively, these efforts have demonstrated the power of combining multiwavelength observations for exoplanet applications, yet few provide complete coverage for stars with directly imageable habitable zones. Despite their strengths, most existing SEDs do not fully cover the parameter space of stars likely to be targeted by HWO and the ELTs, limiting their direct applicability to the planning and interpretation of future high-contrast imaging observations. Most of these existing SEDs focus on M and K dwarfs, while the majority of HWO target stars are of spectral types F, G, and K \citep{mamajek2023}. Furthermore, studies have shown that the use of proxy stars can be unreliable for estimating UV and X-ray fluxes \citep{teal2022, Wilson2025a}, underscoring the need for tailored, empirically constrained panchromatic spectra for HWO-relevant targets. Such data are essential for precursor science, including refining coronagraph exposure-time calculations and informing target selection through photochemical modeling of exoplanet atmospheres and estimates of atmospheric escape---key steps toward assessing whether planets in these systems could maintain habitable environments \citep{Peacock2025}.

Despite the importance of these measurements, archival analyses reveal substantial gaps in the high-energy coverage of high-priority targets identified for HWO. The top tier HWO Target Stars and Systems 2025 list \citep[TSS25;][]{Tuchow2025}, which comprises the NASA Exoplanet Exploration Program's Mission Star List for HWO \citep{mamajek2023}, includes 164 stars prioritized for characterization of potentially habitable exoplanet systems. However, existing observations show that the available high-energy data for these stars are highly heterogeneous, with significant variation in both spectral completeness and quality \citep{Harada2024, Peacock2025}. These gaps limit the ability to construct a comprehensive SED library of target stars. In the X-ray regime, only 26\% of the sample has data suitable for spectral modeling with Astrophysical Plasma Emission Code (APEC) models. For many stars, the best available measurements are shallow \textit{ROSAT} detections, and others lack X-ray observations entirely. Restricting to stars with \textit{Hubble Space Telescope} (\textit{HST}) observations, only 31\% have spectroscopy in both the NUV and FUV; just 40\% have any FUV spectroscopy, and only 35\% have FUV photometry. A small fraction of stars have measurements of Ly$\alpha$ or reconstructed EUV fluxes, despite their critical role in driving upper-atmospheric escape.

\begin{figure*}[t!]
    \centering
    \includegraphics[width=0.95\textwidth]{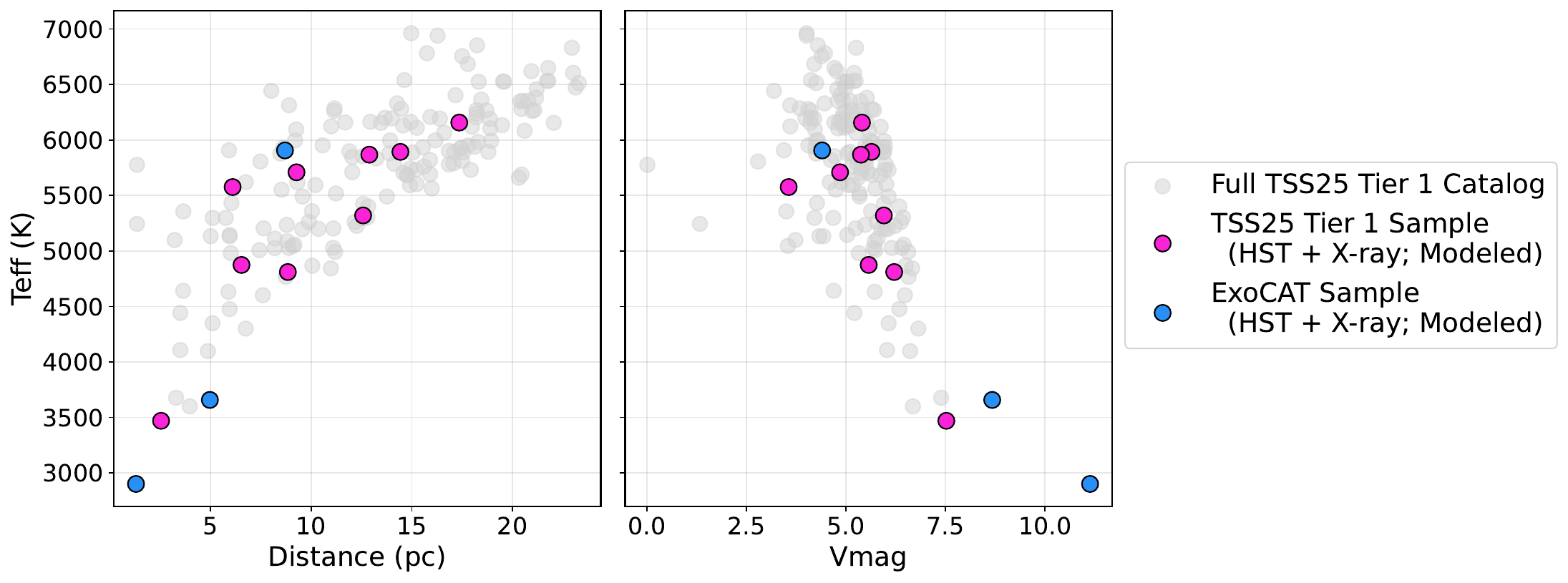}
    \caption{Demographic comparison of our modeled stellar samples with the full TSS25 Tier 1 catalog. Gray points show all Tier 1 stars, while colored symbols highlight the subsets with both HST and X-ray observations that are included in our modeling sample. Magenta points denote stars drawn from the TSS25 Tier 1 list, and blue points denote additional stars from the ExoCAT database with comparable observations. Left: stellar effective temperature versus distance. Right: effective temperature versus apparent V-band magnitude. }
    \label{fig:demo}
\end{figure*}

The scarcity becomes even more pronounced when considering multiwavelength completeness across all high-priority stars. Of the 164 stars, only 51 have any X-ray observations. Among these, 22 are too faint for robust spectral modeling, and eight are unresolved binaries, leaving just 19 stars with reliable X-ray spectra. Within this subset, only nine have both \textit{HST} FUV and NUV coverage.\footnote{At the time of writing, the remaining eight stars lack both FUV and NUV data, while two have existing NUV observations and require only new FUV data. Observations for these ten stars are scheduled as part of \textit{HST} GO 18001.}

To extend this analysis, we also consider likely targets for ELT observations, which tend to favor nearby stars and include both FGK and M dwarfs, with a relatively inclusive treatment of binaries and multiple systems. These stars are drawn from the Nearby Stellar Systems Catalog for Exoplanet Imaging Missions \citep[ExoCAT,][]{Turnbull15}, which overlaps substantially with the HWO target list. \citet{Binder2024} analyzed archival X-ray data for this sample, producing spectra and measuring luminosities and variability for 57 nearby FGKM systems. Cross-matching this sample with the TSS25 list and the \textit{HST} archive yields three additional stars with spectral coverage spanning the X-ray, FUV, and NUV bands.

Here, we present panchromatic SEDs for 12 nearby stars, spanning spectral types F9 to M5.5, each with a directly imageable habitable zone. We combine archival X-ray, FUV, and NUV spectra with photospheric and chromospheric models to produce the most complete view to date of their stellar radiation environments. These SEDs provide a benchmark resource for modeling planetary atmospheres, performing mission simulations, and prioritizing targets for HWO and the ELTs. The paper is organized as follows: Section \ref{sec:sample} describes the stellar sample; Section \ref{sec:assembly} details the archival data and modeling efforts in the construction of the SEDs; Section~\ref{sec:analysis} compares the high energy environments of the 12 stars to the Sun, assesses the validity of the computed EUV emission from each star and discusses implications for the habitability of each system; and Section~\ref{sec:summary} summarizes our conclusions.

\section{Stellar Sample}\label{sec:sample}

Our analysis focuses on nearby FGKM stars from the highest-priority HWO target pool for which sufficient multiwavelength data exist. We identify 12 stars with both robust X-ray and \textit{HST} UV observations, drawn primarily from the TSS25 Tier 1 catalog and supplemented by comparable systems from the ExoCAT database. These objects form the subset of HWO and ELT targets currently best suited for constructing empirical high-energy spectral energy distributions (SEDs). Figure~\ref{fig:demo} places this modeled subset in the demographic context of the full TSS25 Tier 1 catalog. Although this sample represents only a small fraction ($\sim$6\%) of the full catalog, it consists of the currently best-characterized systems with consistent X-ray, FUV, and NUV coverage.

The sample spans more than an order of magnitude in stellar luminosity, from solar analogs to mid-M dwarfs, with effective temperatures ranging from about 2900--6300 K. Table~\ref{tab:stellar_properties} summarizes the fundamental stellar properties adopted in this work. The sample is dominated by solar-type (G stars) and early-K dwarfs, with three M dwarfs extending coverage to cooler temperatures. A noticeable gap appears among mid- to late-K stars ($\sim$3600--4800 K), highlighting an incompleteness in current multiwavelength coverage for intermediate-temperature stars. This regime is particularly important for connecting solar-type and low-mass stellar environments in models of habitable-zone irradiation. The addition of ExoCAT targets modestly broadens the temperature range but does not fully bridge this gap.

In addition to the fundamental parameters listed in Table~\ref{tab:stellar_properties}, we also compile literature measurements of stellar age, activity and rotation diagnostics in Table \ref{tab:activity}. Literature age estimates range from approximately 0.2 to over 12 Gyr, although in several cases different age-dating techniques yield substantially discrepant values. These discrepancies, which can reach several Gyr for individual stars, reflect the intrinsic challenges of stellar age determination for field FGKM stars and motivate our use of multiple activity diagnostics when interpreting trends in high-energy emission.

The sample correspondingly spans a wide range of rotation and activity regimes. Measured rotation periods range from 1.4 days for the rapidly rotating subgiant $\delta$ Pav to 116.6 days for the slowly rotating M dwarf Proxima Centauri, while projected rotational velocities range from effectively 0 km s$^{-1}$ up to nearly 10 km s$^{-1}$. The solar analogs in the sample typically exhibit intermediate rotation periods of 5–10 days, consistent with moderately active stars, while the long rotation periods of the M dwarfs indicate comparatively evolved magnetic states despite their persistent chromospheric activity. Together, these parameters indicate that the sample includes both young, magnetically active stars and older, more weakly active systems.

In addition to rotation and chromospheric activity indicators, Table~\ref{tab:activity} lists published stellar activity-cycle periods where available. Several stars in our sample exhibit reported long-term activity cycles, including $\iota$ Hor, $\kappa^1$ Cet, GL~892, 55 Cnc A, GL~183, and Proxima Centauri. Activity cycles have also been reported for $\chi^1$ Ori and GL~311, although the periods are less well constrained. For Lalande 21185, long-term variability on a $\sim$7 yr timescale has recently been reported \citep{Oviedo2026}, although this has not yet been established as a periodic activity cycle.


\input{stellar_props_REVISED}

\begin{deluxetable*}{l l l l l l}
\tablecaption{Stellar Activity and Rotation Properties}\label{tab:activity}
\tablehead{
\colhead{Name} &
\colhead{SpT} &
\colhead{Age (Gyr)} &
\colhead{$v\sin i$ (km s$^{-1}$)} &
\colhead{$P_{\rm rot}$ (days)} &
\colhead{Activity Cycle Period (years)}
}
\startdata
iota Hor & F9V &$0.47; 2.72^{a,b}$ &$5.53\pm0.39^{c}$ &$8.22^{d}$ & 1.6$^{r,s}$\\
$\chi^1$ Ori & G0V &$0.35; 4.32^{b,f}$ &$8.9\pm0.5^{g}$ &$5.36^{h}$ & $\sim$2, 6, or 8$^{t}$\\
GL311 ($\pi^1$ UMa) & G1.5V &$0.2^{e}$ &$9.6\pm0.5^{g}$ &$5.22^{h}$ &$\sim$13 and $\sim$2.1$^{u}$\\
GL620-1A & G1V &$2;<1;0.4^{b,f,i}$ &$2.07\pm0.98^{j}$ &$6.10^{d}$ &\nodata\\
$\kappa^1$ Cet & G5V &$0.35;0.7;2.2^{b,e,f}$ &$6.1\pm1.1^{k}$ &$9.08^{l}$ &$\sim$6 and 3.1$^{v,w}$\\
$\delta$ Pav & G8IV &$6.2^{b}$ &$1.51\pm0.19^{j}$ &$1.41^{d}$&\nodata \\
55 Cnc A ($\rho^1$ Cnc) & K0IV/V &$9.5^{b}$ &$2.3\pm0.64^{m}$ &$37.4^{h}$ &$\sim$12.6$^{x}$\\
GL 892 & K3V &$11.0\pm2.2,12.46^{n,b}$ &$1.64\pm0.59^{m}$ &$11.60^{d}$ &2.95 and 12.7$^{y}$\\
GL 183 & K3V &$2,5.3,<5.45^{e,b,f}$ &$3.1\pm1.1^{k}$ &$39.3^{l}$&11.1$^{v}$ \\
GJ 832 & M1.5V &$9.24^{b}$ &\nodata &$45.7^{o}$ &\nodata\\
Lalande 21185 & M2V &$8.05^{+3.96,q}_{-4.52}$; 10$^{b}$ &$3.0\pm0.98^{p}$ &56.16$\pm0.27^{q}$ &$\sim$4.9$^{z}$\\
Proxima Centauri (GJ 551) & M5.5V &$4.85^{b}$ &\nodata &$116.6^{o}$ &$\sim$7$^{aa}$\\
\enddata
\tablecomments{
{\bf References:}
$^{a}$ \citealt{SanzForcada2010};
$^{b}$ \citealt{Turnbull15};
$^{c}$ \citealt{Soto2018};
$^{d}$ \citealt{Fetherolf2023};
$^{e}$ \citealt{Mamajek2008};
$^{f}$ \citealt{Takeda2007};
$^{g}$ \citealt{Brewer2016};
$^{h}$ \citealt{Mittag2017};
$^{i}$ \citealt{Ghezzi2010};
$^{j}$ \citealt{Costa2020};
$^{k}$ \citealt{Luck2017};
$^{l}$ \citealt{Olah2016};
$^{m}$ \citealt{Rainer2023};
$^{n}$ \citealt{Gillon2017};
$^{o}$ \citealt{Suarez2015};
$^{p}$ \citealt{Holmberg2007};
$^{q}$ \citealt{Hurt2022};
$^{r}$\citealt{Sanz-Forcada2013};
$^{s}$\citealt{Sanz-Forcada2019};
$^{t}$\citealt{rosen2016};
$^{u}$\citealt{Messina2002};
$^{v}$\citealt{Baliunas1995};
$^{w}$\citealt{BoroSaikia2022};
$^{x}$\citealt{Baluev2015};
$^{y}$\citealt{Metcalfe2013};
$^{z}$\citealt{Oviedo2026};
$^{aa}$\citealt{Wargelin2017}.
}
\end{deluxetable*}

\section{Assembly of Panchromatic Spectral Energy Distributions}\label{sec:assembly}

\subsection{Archival Data}

\subsubsection{X-ray}\label{subsec:Xray}
X-ray observations were obtained from the XMM-Newton and Chandra archives. Every star in the sample has XMM-Newton observations, and two targets ($\kappa^1$ Ceti and Proxima Centauri) have detections with both Chandra and XMM-Newton. All targets except Lalande 21185 were analyzed in detail by \citet{Binder2024}, who performed uniform data reprocessing, astrometric cross-matching, spectral modeling using APEC models, and data quality assessment. The XMM-Newton observations of Lalande 21185 were processed and modeled following the same procedures (Schwieterman et al., in prep.). For all stars, we adopt the X-ray spectra provided by \citet{Binder2024} and Schwieterman et al. (in prep.), specifically using the \texttt{quiescent\_lines} data products, which exclude time intervals affected by elevated or flaring emission.

For stars with multiple X-ray observations, we adopt the averaged quiescent-state spectra from \citet{Binder2024}, which were corrected for instrument response and combined into representative spectra. These spectra were then extrapolated over a broader energy range for SED construction. For the FGK stars, the best-fit quiescent spectral models show minimal variability between epochs despite multi-year baselines.

Lalande 21185 is a notable exception. Two quiescent spectra are available (from 2001 and 2004), which differ by approximately a factor of 27 in flux ($\log F_{X,2001} = -13.38$ erg s$^{-1}$ cm$^{-2}$ and $\log F_{X,2004} = -11.94$ erg s$^{-1}$ cm$^{-2}$). To account for this variability, we construct two SEDs for this star using each X-ray spectrum. Throughout the analysis we report ranges that encompass these differences; unless otherwise noted, figures show the SED constructed using the 2001 X-ray spectrum.


\subsubsection{Ultraviolet}\label{subsec:UV}
UV spectra for all targets were retrieved from the Mikulski Archive for Space Telescopes (MAST) and include observations obtained with the \textit{Hubble Space Telescope} (\textit{HST}) Space Telescope Imaging Spectrograph (STIS) and/or the Cosmic Origins Spectrograph (COS). All stars in the sample have coverage in both the FUV and NUV wavelength ranges. The majority of the data come from the high-resolution STIS echelle modes (E140M/H and E230M/H), supplemented by STIS G140M, G230L, and G230LB modes and COS G130M, G160M, and G230L spectra. The available observations, including instrumental configurations and observation dates, are summarized in Table~\ref{tab:obs} and can be accessed via \dataset[doi:10.17909/89p4-zx80]{https://doi.org/10.17909/89p4-zx80}.

For targets with multiple observations obtained in similar instrumental configurations, we preferentially adopted the most recent datasets, which typically provide improved signal-to-noise and more complete wavelength coverage. When contemporaneous observations across multiple wavelength bands were available, those datasets were preferentially selected to minimize the effects of stellar variability.

All observations were retrieved following standard processing with the CALSTIS and CALCOS pipelines using the most recent calibration reference files. For STIS echelle observations, we adopted the HASP CSPEC coadded spectral products.\footnote{\url{https://www.stsci.edu/files/live/sites/www/files/home/hst/instrumentation/cos/documentation/instrument-science-reports-isrs/_documents/ISR2024-01.pdf}} When available, we adopted High-Level Science Products (HLSPs) from the StarCAT \citep{Ayres2010} and MUSCLES \citep{France2016,Youngblood16,Loyd2016} catalogs in place of standard pipeline products. These HLSPs provide coadded, wavelength-aligned spectra with improved background subtraction and flux consistency across echelle orders and exposures. For Lalande 21185, we used the G230LB spectrum from the STIS Next Generation Spectral Library HLSP \citep{Gregg2004}, which incorporates scattered red-light correction.

All spectra were visually inspected to assess data quality. Regions affected by low-quality pixels or calibration artifacts, particularly near the blue wing of Ly$\alpha$ ($\lambda < 1205$~\AA), were masked. For targets with multiple visits, exposures were aligned in wavelength space prior to coaddition. Geocoronal emission features, including the core of Ly$\alpha$, \ion{O}{1} 1304~\AA, and \ion{O}{1} 1356~\AA, were identified and masked.

For most targets, the FUV and NUV observations were obtained contemporaneously within the same \textit{HST} visit or program. However, five stars (GL~311, 55~Cnc~A, GL~892, Lalande~21185, and Proxima Centauri) have FUV and NUV observations separated by more than two years, with Lalande~21185 and Proxima Centauri showing the largest separations of more than 17 years.

Two of these stars, 55~Cnc~A and GL~892, are older early-K dwarfs ($\gtrsim$6 Gyr) for which relatively low levels of long-term UV variability are expected. GL~311 is a younger K dwarf (0.2 Gyr; \citealt{Mamajek2008}) and may exhibit larger intrinsic variability; however, the NUV spectra of solar-type stars are dominated by photospheric emission, which reduces the impact of activity-related variability on our analysis.

Lalande~21185 (a $\sim$10 Gyr M2V star; \citealt{Turnbull15}) represents a special case due to both the long temporal baseline and the availability of multiple NUV datasets. Lalande~21185 has two available NUV observations: a STIS G230LB spectrum obtained in 2002 and a STIS E230H spectrum obtained in 2019--2020 contemporaneously with the FUV E140M observations. The G230LB spectrum provides substantially broader wavelength coverage ($\sim$2200--10110~\AA) compared to the E230H spectrum ($\sim$2570--2840~\AA), which is optimized for high-resolution measurements of the \ion{Mg}{2} h\&k lines. Comparison of the \ion{Mg}{2} fluxes between these datasets shows agreement at the $\sim$10\% level ($\log F = -13.351$ and $-13.396$ erg s$^{-1}$ cm$^{-2}$ for the G230LB and E230H observations, respectively). We therefore adopt the broader-coverage G230LB spectrum for SED construction. This choice could introduce a small systematic uncertainty in the FUV/NUV flux ratio at the $\sim$10\% level, but does not affect the overall conclusions of this work.

\input{table_observations}

\subsection{Observational Epoch Comparison}

\input{obs_epoch_table}

Because the X-ray and UV observations were obtained over a wide range of epochs, we summarize the observation dates and their temporal separations in Table~\ref{tab:epoch_summary}. For stars with extensive X-ray monitoring, we list the observation date ranges and number of observations rather than individual epochs. The UV and X-ray data are typically separated by months to decades, which may introduce variability associated with stellar activity cycles.

Several targets have observations that are nearly contemporaneous across wavelength regimes. For example, GL~620-1A and GJ~832 have UV and X-ray observations obtained within days of each other, minimizing uncertainties associated with long-term variability. Similarly, $\chi^1$ Ori, GL~183, and $\delta$ Pav have observations separated by less than one year.

In contrast, some targets show substantial temporal separations between wavelength regimes. GL~311 has UV observations obtained roughly a decade after the available X-ray spectrum, while Lalande 21185 and Proxima Centauri have separations approaching two decades between the earliest and most recent observations. $\kappa^1$ Cet also shows a wide temporal baseline due to the availability of both early XMM-Newton observations and more recent XMM-Newton and Chandra monitoring.

For stars with multiple X-ray observations spanning many years (e.g., $\iota$ Hor, $\kappa^1$ Cet, and Proxima Centauri), the adopted spectra represent averages of quiescent emission states as described in Section \ref{subsec:Xray}. This approach is intended to provide representative coronal emission levels rather than measurements tied to any single epoch.

The published activity-cycle periods (Table \ref{tab:activity}) provide additional context for assessing the impact of these temporal separations. For several targets, the UV--X-ray separation is short compared with the reported activity-cycle period (e.g., $\chi^1$ Ori and GL~183), suggesting that differences in activity phase are likely to be relatively small. For $\iota$ Hor, the $\sim$1.6 yr activity cycle is shorter than the 2011--2018 X-ray monitoring baseline, such that the adopted X-ray spectrum averages over multiple activity cycles, while the UV observations represent a single epoch. Conversely, the UV and X-ray observations of 55 Cnc A, GL~892, Lalande 21185, and Proxima Centauri span timescales comparable to or longer than their reported activity cycles, making differences in activity phase a potentially important source of uncertainty. The same may apply to $\kappa^1$ Cet, whose UV observations predate the more recent X-ray monitoring by nearly two decades, although the adopted X-ray spectrum incorporates observations spanning multiple epochs. For GL~311, the UV and X-ray observations are separated by approximately one activity-cycle period, although the reported cycle is less securely established.

When multiple UV observations covering the same wavelength band were available, we compared the observations to assess whether the adopted dataset was anomalous relative to other epochs. The available observations were generally consistent within the observed level of variability, and we found no cases in which the adopted, most-recent observation was a clear outlier in flux or activity level. In a small number of cases, multiple observations were associated with target-acquisition failures in the initial observations (e.g., $\iota$ Hor and GL~620-1A), rendering those data unusable. For the targets with multiple usable UV observations, the selected observation was either closer in time to the X-ray data, fell within the range of X-ray epochs used to construct the averaged APEC spectrum, or was separated from the X-ray data by more than $\sim$10 years along with the other available UV observation, such that the choice of UV epoch did not provide a meaningful improvement in temporal correspondence. Thus, the adopted UV spectra were not selected over more contemporaneous data on the basis of an unexamined difference in activity level.

We do not attempt to correct the SEDs for stellar activity cycle phase or long-term variability. Although published activity cycles provide characteristic timescales for several stars in our sample, the available observations are too heterogeneous to determine the activity phase consistently across all wavelength regimes, and the X-ray spectra themselves represent averages over multiple epochs for several targets. The resulting SEDs should therefore be interpreted as representative, time-averaged stellar emission distributions rather than strictly simultaneous measurements. For the most active stars and for targets with UV--X-ray separations comparable to or exceeding their reported activity-cycle periods, activity-cycle variability represents an additional source of systematic uncertainty when interpreting the relative flux levels across wavelength regimes.

\subsection{PHOENIX Upper Atmosphere Models}

We computed synthetic spectra ($>$100 \AA) of each star using the most recent version of the \texttt{PHOENIX} stellar atmosphere code (V20.01) \citep{Hauschildt1993, Hauschildt2006, Baron2007}, following the methodology of \citet{Peacock2022}. These models were guided by the archival HST spectra and were used to fill gaps in the FUV and NUV wavelength coverage, predict the unobservable extreme ultraviolet (EUV) spectrum, and extend the SEDs to wavelengths redward of the available UV observations.

The \texttt{PHOENIX} models include semi-empirical upper atmospheres with prescribed chromospheric and transition region temperature structures extending to temperatures of 200,000 K. This temperature range encompasses the formation regions of the observed FUV and NUV emission lines but does not include the coronal plasma ($\sim10^6$ K) responsible for X-ray emission. These upper atmospheres are connected to an underlying photosphere in radiative–convective equilibrium defined by the stellar effective temperature ($T_{\rm eff}$), surface gravity (log $g$), stellar mass ($M_\star$), and metallicity ([Fe/H]) adopted for each target (Table \ref{tab:stellar_properties}).

For each star we computed a grid of 72 upper atmosphere models by varying three free parameters describing the chromosphere and transition region structure: (1) the column mass at the base of the chromosphere, (2) the column mass at the top of the chromosphere, and (3) the temperature gradient through the transition region.

From these grids we identified a representative model by minimizing the reduced $\chi^2$ statistic computed from comparisons between model and observed emission surface line fluxes (Table \ref{tab:linefluxes}). The fitting procedure primarily relied on strong transition region and chromospheric diagnostics including \ion{H}{1} Ly$\alpha$, \ion{Si}{2}, \ion{N}{5}, \ion{C}{2}, \ion{C}{4}, \ion{Si}{4}, \ion{Al}{2} and \ion{Mg}{2} (where available for each star). For doublet lines (e.g., \ion{N}{5}, \ion{Si}{4}, and \ion{Mg}{2}), fluxes were measured by integrating each component separately over windows centered on the line cores and summing the resulting fluxes.

Because interstellar absorption and geocoronal contamination affect the Ly$\alpha$ line core, the model selection primarily used the wings of the observed profile where reliable flux measurements could be obtained. For consistency across the sample, the final SEDs adopt the \texttt{PHOENIX} Ly$\alpha$ profiles rather than performing individual line reconstructions, with the exception of GJ 832 and Proxima Centauri, where previously published MUSCLES reconstructions were adopted \citep{Youngblood16}.

\subsection{Final SED Assembly}

\begin{figure*}
    \centering
    \includegraphics[width=0.49\linewidth]{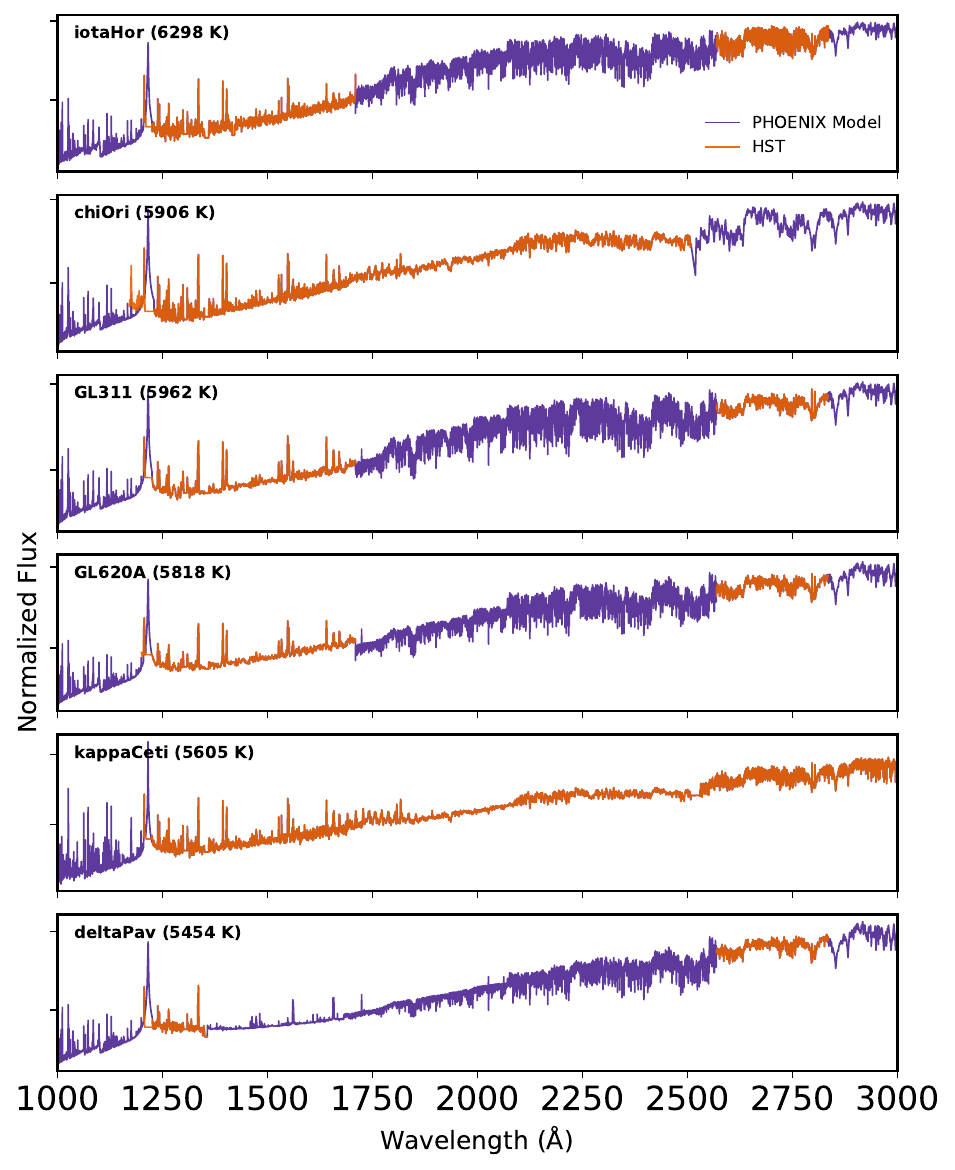}
    \includegraphics[width=0.49\linewidth]{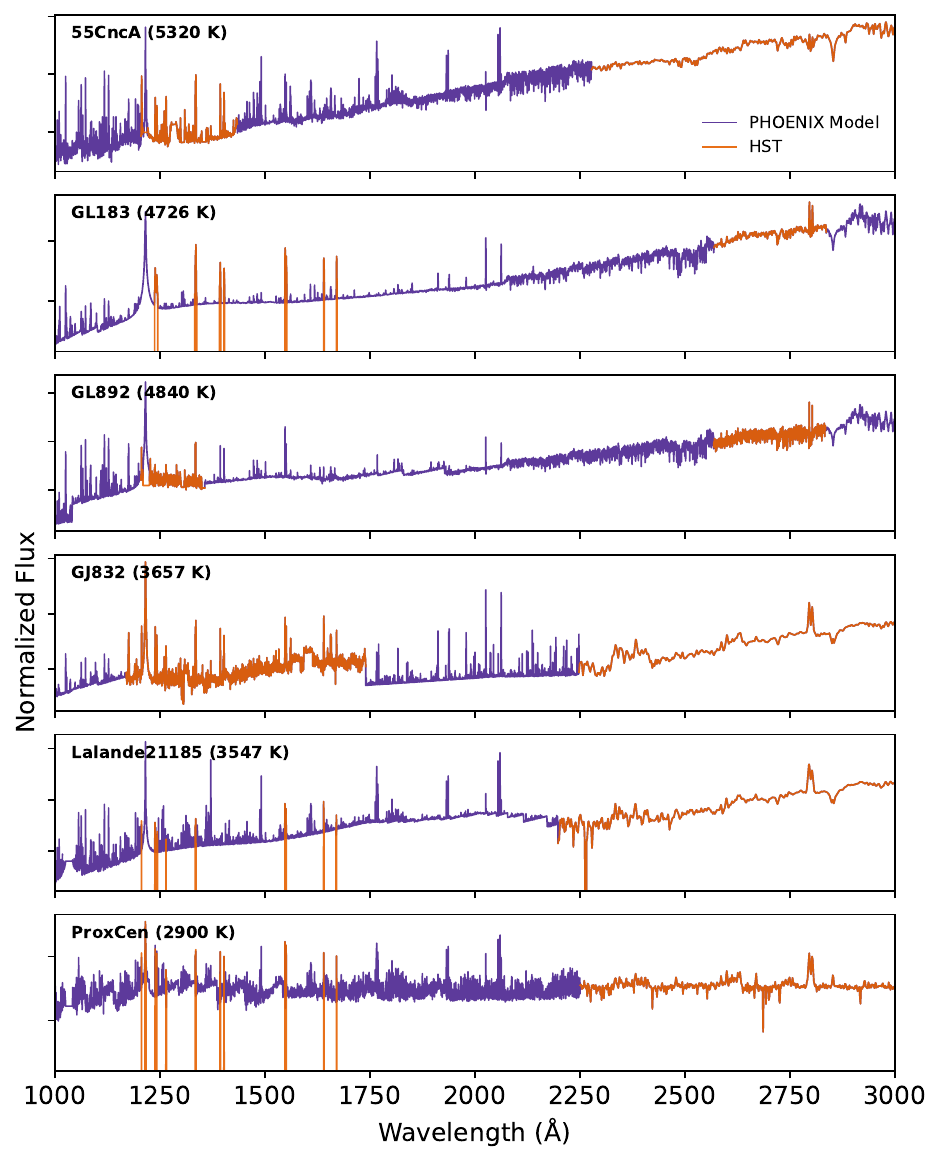}
    \caption{Ultraviolet spectra of the sample stars, ordered by effective temperature. The HST observations adopted for the final SEDs are shown in orange, while the corresponding \texttt{PHOENIX} model spectra are shown in purple. The models provide wavelength coverage where observations are unavailable, including the unobservable EUV region and wavelengths longward of the NUV HST spectra. For GL 183, Lalande 21185, and Proxima Centauri, FUV emission lines are displayed with extended baselines below the continuum to aid visualization; the final SED product instead smoothly merges these features with the surrounding continuum.}
    \label{fig:UVstacks}
\end{figure*}

The final panchromatic SEDs were constructed by combining the \texttt{PHOENIX} model spectra with available observations across multiple wavelength regimes. In the UV, the \texttt{PHOENIX} spectra were combined with the \textit{HST} observations following the procedures described in Section \ref{subsec:UV} (only observed data longward of 1205 \AA\ were used, and geocoronal emission lines were masked). The \texttt{PHOENIX} spectra were then inserted to fill gaps between observational bandpasses. The combined UV spectra are shown in Figure \ref{fig:UVstacks}. The complete SEDs are publicly available through Zenodo\footnote{\dataset[10.5281/zenodo.22001080]{https://doi.org/10.5281/zenodo.22001080}
}. The SEDs are provided as machine-readable FITS files containing wavelength and flux density and are available in both native and 1 \AA\ resolution.

The reliability of continuum measurements varies across the sample. For the M dwarf targets, the FUV continuum is detected in GJ 832 at 6.1$\sigma$ confidence \citep{Loyd2016}, and the observed spectrum was therefore retained. In contrast, the FUV continuum is not clearly detected in Lalande 21185 or Proxima Centauri, neither is it detected in the K dwarf GL 183. For these stars, only clearly detected emission lines were retained from the observations while the surrounding continuum was supplemented by the \texttt{PHOENIX} model.

At XUV energies, the \texttt{PHOENIX} models were supplemented with coronal emission models, since \texttt{PHOENIX} does not include X-ray temperatures. For all stars, APEC thermal plasma models from \citealt{Binder2024} were used to provide the emission over the 5--180 \AA\ wavelength range. These models provide the high-energy extension needed to complete the panchromatic SEDs. The resulting SEDs therefore combine observational constraints where available with physically motivated models to produce continuous spectra spanning the X-ray through radio wavelength ranges.

\begin{deluxetable*}{lccccccccc}
\tabletypesize{\scriptsize}
\label{tab:linefluxes}
\tablecaption{Integrated UV emission surface line fluxes, continuum subtracted. Fluxes are given in units of $10^{x}$ erg s$^{-1}$ cm$^{-2}$ as indicated in each column.}
\tablehead{
\colhead{Star} & 
\colhead{\shortstack{Si III\\1206 $\AA$\\(F$\times 10^{4}$)}} & 
\colhead{\shortstack{Ly$\alpha$\\1216 $\AA$\\(F$\times 10^{6}$)}} & 
\colhead{\shortstack{N V\\1240 $\AA$\\(F$\times 10^{3}$)}} & 
\colhead{\shortstack{Si II\\1265 $\AA$\\(F$\times 10^{3}$)}} & 
\colhead{\shortstack{C II\\1335 $\AA$\\(F$\times 10^{4}$)}} & 
\colhead{\shortstack{Si IV\\1398 $\AA$\\(F$\times 10^{4}$)}} & 
\colhead{\shortstack{C IV\\1550 $\AA$\\(F$\times 10^{4}$)}} & 
\colhead{\shortstack{Al II\\1671 $\AA$\\(F$\times 10^{3}$)}} & 
\colhead{\shortstack{Mg II\\2800 $\AA$\\(F$\times 10^{6}$)}}
}
\startdata
$\iota$ Hor & $2.91\pm0.12$ & $1.33^{\mathrm{a}}$ & $5.59\pm0.71$ & $2.52\pm0.40$ & $3.15\pm0.09$ & $2.54\pm0.11$ & $3.60\pm0.19$ & $1.22\pm4.55$ & $1.42\pm0.13$ \\
$\chi$ Ori & $6.04\pm0.09$ & $3.70^{\mathrm{a}}$ & $8.43\pm0.30$ & $5.07\pm0.20$ & $6.27\pm0.04$ & $5.31\pm0.05$ & $7.46\pm0.08$ & $25.7\pm4.3$ & \nodata \\
GL 311 & $8.28^{\mathrm{a}}$ & $4.43^{\mathrm{a}}$ & $15.5\pm0.6$ & $6.71\pm0.38$ & $10.1\pm0.1$ & $8.19\pm0.11$ & $13.6\pm0.2$ & $25.6\pm3.0$ & $3.24\pm0.02$ \\
GL 620A & $5.54\pm0.11$ & $1.39^{\mathrm{a}}$ & $6.61\pm0.43$ & $4.26\pm0.28$ & $4.94\pm0.06$ & $4.14\pm0.08$ & $5.47\pm0.12$ & $17.9\pm3.9$ & $1.16\pm0.01$ \\
$\kappa$ Cet & $2.03^{\mathrm{a}}$ & $3.73^{\mathrm{a}}$ & $3.96\pm0.16$ & $2.02\pm0.10$ & $2.14\pm0.02$ & $1.77\pm0.02$ & $2.66\pm0.04$ & $8.60\pm2.18$ & $0.654\pm0.006$ \\
$\delta$ Pav & $0.309^{\mathrm{a}}$ & $0.515^{\mathrm{a}}$ & $0.708\pm0.087$ & $0.629\pm0.015$ & $0.610\pm0.009$ & \nodata & $0.00911^{\mathrm{a}}$ & $0.00299^{\mathrm{a}}$ & $0.295\pm0.002$ \\
55 Cnc A & $0.442^{\mathrm{a}}$ & $0.153^{\mathrm{a}}$ & $1.22\pm0.02$ & $0.886\pm0.018$ & $0.788\pm0.005$ & $0.343\pm0.004$ & $0.358^{\mathrm{a}}$ & \nodata & $0.374\pm0.001$ \\
GL 892 & $0.0543^{\mathrm{a}}$ & $2.75^{\mathrm{a}}$ & \nodata & $0.293\pm0.120$ & $0.370\pm0.007$ & $0.345^{\mathrm{a}}$ & $1.52^{\mathrm{a}}$ & $1.83^{\mathrm{a}}$ & $0.301\pm0.003$ \\
GL 183 & \nodata & $0.156^{\mathrm{a}}$ & $1.27\pm0.20$ & $0.0148^{\mathrm{a}}$ & $0.735\pm0.019$ & $0.257\pm0.295$ & $0.814\pm0.049$ & $2.24\pm6.65$ & $0.452\pm0.002$ \\
GJ 832 & $0.0522\pm0.005$ & $0.190$ & $0.691\pm0.006$ & $0.100\pm0.035$ & $0.123\pm0.001$ & $0.0663\pm0.001$ & $0.160\pm0.002$ & $0.295\pm0.019$ & $0.0251\pm0.0001$ \\
Lalande 21185 & $0.00332^{\mathrm{a}}$ & $0.0191^{\mathrm{a}}$ & \nodata & $0.00464\pm0.00095$ & $0.00178\pm0.00095$ & \nodata & $0.0136\pm0.0039$ & $0.141\pm0.092$ & $0.0184$ \\
Proxima Centauri & $0.112^{\mathrm{a}}$ & $0.750$ & $3.49\pm0.08$ & $0.138\pm0.032$ & $0.626\pm0.006$ & $0.236\pm0.008$ & $1.74\pm0.02$ & $1.63\pm0.12$ & $0.0359\pm0.007$ 
\enddata
\tablenotetext{a}{Value derived from the \texttt{PHOENIX} model rather than direct observation.}
\end{deluxetable*}

\section{Results and Analysis}\label{sec:analysis}

\subsection{High-Energy Radiation Environment Compared to the Sun}

\begin{figure*}
    \centering
    \includegraphics[width=0.75\linewidth]{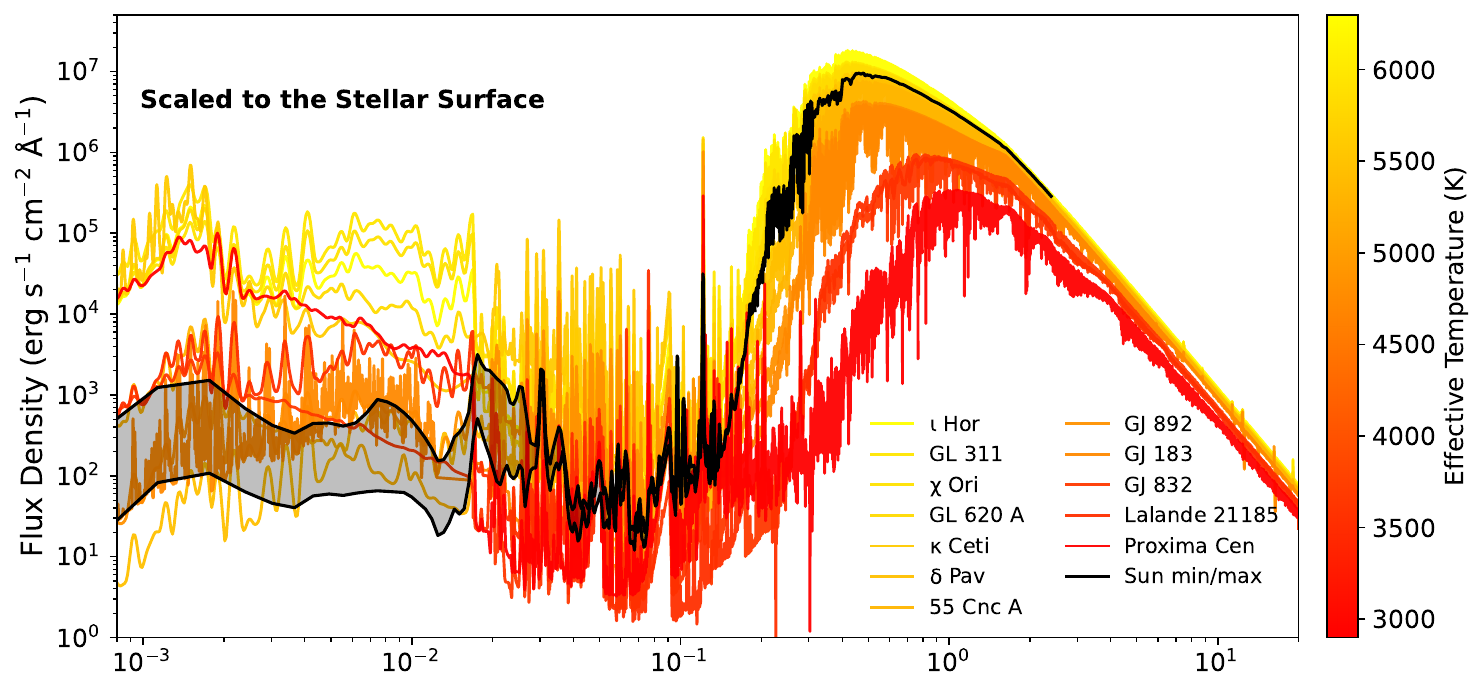}
    \includegraphics[width=0.75\linewidth]{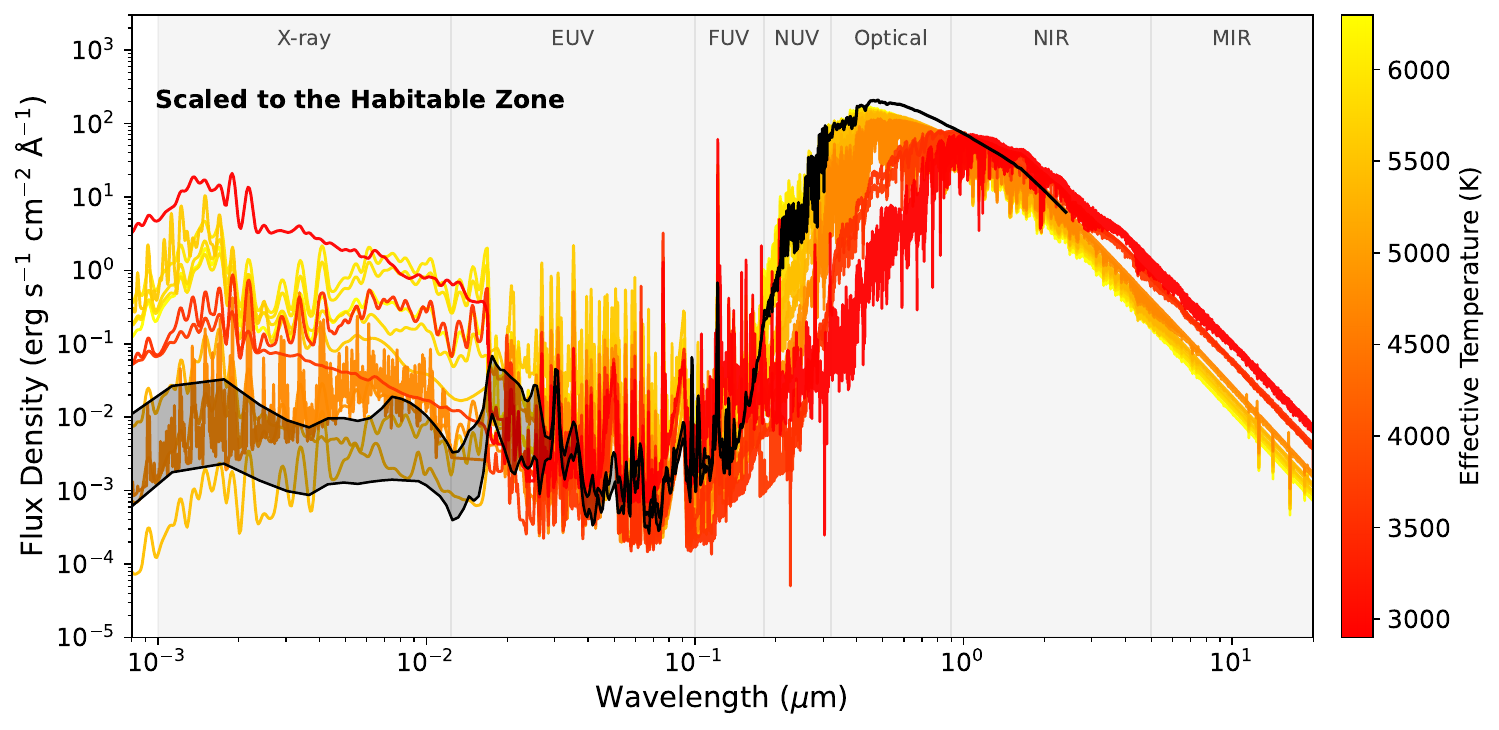}
    \caption{Panchromatic SEDs of the sample stars compared to the Sun. Stellar SEDs are color–coded by effective temperature, while the solar spectrum is shown in black. The shaded region indicates the range of solar fluxes observed over the solar activity cycle. Solar reference spectra are derived from measurements obtained by the Thermosphere Ionosphere Mesosphere Energetics and Dynamics (TIMED) Solar EUV Experiment (SEE) \citep{woods2005,LASP2005}. \textit{Top panel:} flux densities scaled to the stellar surface to enable direct comparison of intrinsic high-energy emission levels. \textit{Bottom panel:} flux densities scaled to the habitable zone distance of each system, illustrating the high-energy radiation environments experienced by potentially habitable planets. The majority of the sample stars exhibit XUV flux levels that exceed those of the Sun, even when compared to solar maximum conditions, indicating elevated high-energy radiation environments relative to the present-day Sun.}
    \label{fig:SED_solar_comp}
\end{figure*}

We compare the stellar SEDs to one another and to the Sun in Figure \ref{fig:SED_solar_comp}, both scaled to the stellar surface and to the respective HZs of each system. The HZ fluxes are computed following \cite{Kopparapu2014}, assuming a 1 $M_\oplus$ planet located at the midpoint between the Recent Venus and Early Mars limits. We find that most stars in this sample emit XUV fluxes exceeding solar values, even when compared to the Sun at activity maximum, both at the stellar surface and when scaled to their HZ distances.

\begin{figure}
    \centering
    \includegraphics[width=0.5\linewidth]{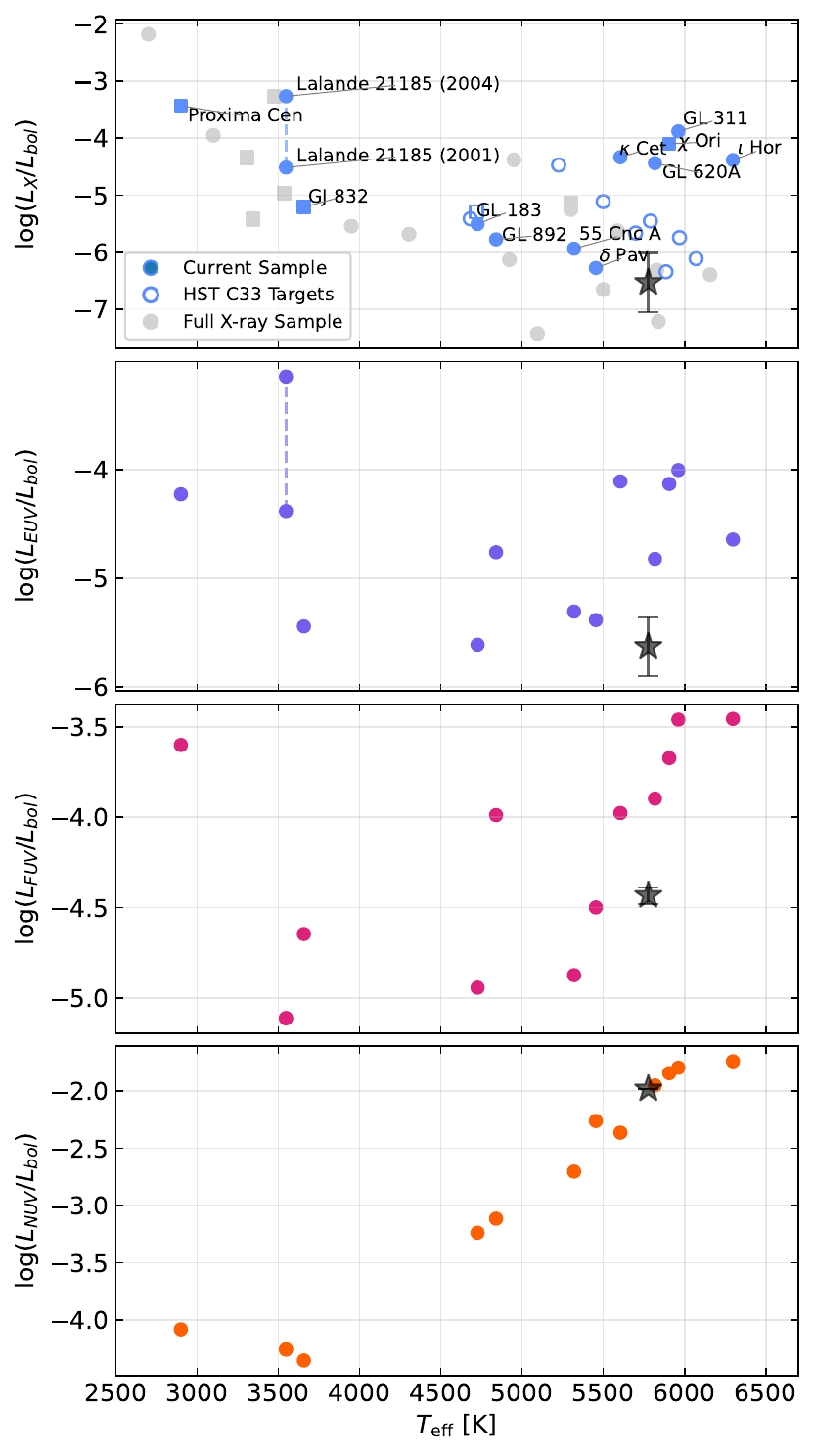}
    \caption{Fractional high-energy luminosities of the sample stars as a function of effective temperature. The Sun is shown as a black star symbol, with vertical error bars indicating the range of solar emission over the activity cycle. In the top panel, gray points show the full comparison sample of stars with X-ray measurements from \citealt{Binder2024}. Circles denote targets included in the TSS25 sample, while squares indicate stars present only in ExoCat. Open symbols indicate targets with planned FUV and/or NUV observations in HST Cycle 33. The majority of the current sample exhibits fractional X-ray and EUV luminosities approximately 1-2 orders of magnitude higher than solar values, although a small number of targets are consistent with solar levels. The FUV luminosities of solar-type stars are generally comparable to, but slightly elevated relative to, solar values. For Lalande 21185, two quiescent APEC models from Schwieterman et al. (in prep) are used to construct separate SEDs, which are reflected in the X-ray and EUV panels. More broadly, the sample exhibits measurable X-ray variability \citep{Binder2024}, with Proxima Centauri and GJ 832 varying by approximately an order of magnitude (comparable to Lalande 21185) while the remaining targets show more modest variability at the $\sim$0.5 dex level.}
    \label{fig:luminosity_scatter}
\end{figure}

In Table \ref{tab:luminosities} and Figure \ref{fig:luminosity_scatter}, we present the X-ray, EUV, FUV, and NUV luminosities of the sample alongside the solar range across its activity cycle. Figure \ref{fig:luminosity_scatter} shows these values as fractional luminosities ($L_{\rm band}/L_{\rm bol}$) as a function of effective temperature. In the X-ray panel we also include the 57 stars with X-ray measurements from \citealt{Binder2024}, who found that 28 systems have $L_{\rm X}/L_{\rm bol}$ ratios comparable to or below that of the modern or young Sun. Within our sample, only three stars ($\delta$ Pav, 55 Cnc A, and GL 183) fall clearly within the modern solar range, although three additional stars scheduled for HST Cycle 33 UV observations also show similarly low activity levels.

The EUV luminosity ratios closely track the X-ray behavior, with $\delta$ Pav and 55 Cnc A again appearing most solar-like. GL 183 and GJ 832 also show EUV luminosity ratios comparable to solar values, although their X-ray luminosity ratios are slightly elevated relative to the Sun. This agreement is notable because much of the EUV wavelength range (180--912 \AA) is computed from \texttt{PHOENIX} models constrained primarily by FUV and NUV emission lines rather than X-ray observations, although the contribution of the APEC component at shorter EUV wavelengths varies substantially among the stars. The consistency of the EUV and X-ray trends therefore provides confidence that the semi-empirical EUV reconstructions capture the expected activity scaling.

Although a substantial fraction of the integrated EUV luminosity generally arises from the longer-wavelength ($>180$ \AA) portion of the spectrum, the contribution from the coronal APEC component is significant for some stars. The APEC component contributes 4--75\% of the total integrated 100--912 \AA\ EUV flux across the sample, with a median contribution of 37\% (Table \ref{tab:luminosities}). The largest APEC contributions occur for $\iota$ Hor, $\chi^1$ Ori, GL~311, Lalande 21185, and Proxima Centauri, for which the coronal component contributes 58--75\% (or 98\% in the case of the SED for Lalande 21185 that adopts the 2004 X-ray observation) of the total EUV flux. In contrast, the APEC contribution is $\leq$14\% for $\delta$ Pav, $\kappa^1$ Cet, 55 Cnc A, and GL~892. This variation is important when interpreting EUV scaling relations (Section \ref{subsec:euvcomparisons}), as relationships involving shorter wavelength ranges can be more sensitive to the coronal APEC component for stars with large APEC contributions. 

At longer wavelengths, the FUV luminosities show that most solar-type stars in the sample exceed solar fractional luminosities by roughly 0.5–1 dex, with $\delta$ Pav and 55 Cnc A again standing out as the most solar-like cases. A general decrease in FUV luminosity with decreasing effective temperature is observed among the G and K stars, although the M dwarfs show significant scatter, reflecting the diversity of magnetic activity levels at low masses. The NUV luminosities show the clearest temperature dependence, with a monotonic decrease toward cooler stars and the Sun falling naturally along this trend.

GL 892 stands out as a notable outlier, with elevated FUV emission that propagates into a relatively high EUV luminosity compared to stars of similar temperature. This is somewhat unexpected given its estimated age of 11–12 Gyr (Table \ref{tab:activity}), modest projected rotation velocity ($v\sin i = 1.64$ km s$^{-1}$), and rotation period of 11.6 days. However, the available UV constraints for this star are more limited than for the rest of the sample, with only three emission lines (\ion{Si}{2}, \ion{C}{2}, and \ion{Mg}{2}) available to guide the \texttt{PHOENIX} chromospheric structure. This limited observational input may contribute to the discrepancy. We revisit this issue in Section \ref{subsub:scalings} and Table \ref{tab:euvcomps}, where we compare our EUV estimates to empirical scaling relationships based primarily on X-ray fluxes.

\begin{figure}
    \centering
    \includegraphics[width=0.5\linewidth]{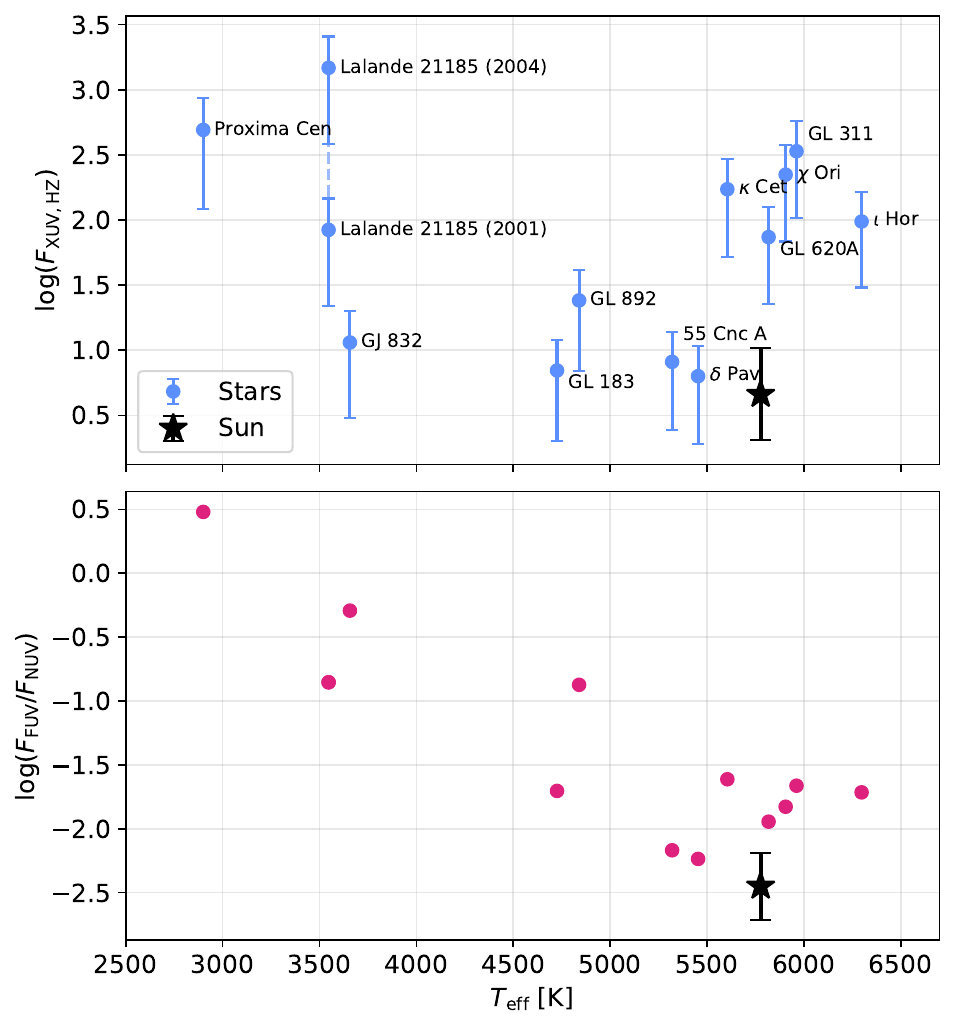}
    \caption{High-energy radiation environments of the sample stars as a function of effective temperature. The top panel shows the XUV flux (log($F_{\rm XUV}$)) within the habitable zone, while the bottom panel shows the ratio of FUV to NUV flux (log($F_{\rm FUV/NUV}$)). The Sun is shown as a black star for reference. Vertical error bars on the sample stars indicate the range of fluxes between the inner and outer habitable zone boundaries and do not represent measurement uncertainties, while the solar error bars reflect both the habitable zone width and the range of solar emission across the activity cycle. Consistent with Figure \ref{fig:luminosity_scatter}, most stars in the sample exhibit habitable zone XUV fluxes approximately 1-2 orders of magnitude higher than those experienced by the Earth, and the FUV to NUV flux ratio of solar-temperature stars are slightly elevated relative to solar values, with even more relatively stronger FUV emission seen toward later spectral types.}
    \label{fig:HZ_fluxes}
\end{figure}

In Figure \ref{fig:HZ_fluxes}, we compare the XUV fluxes throughout the HZ and the FUV/NUV flux ratios to the present-day solar environment at Earth. The XUV flux is a key driver of atmospheric heating and escape, while the FUV/NUV ratio strongly influences atmospheric photochemistry. We find that most stars produce HZ XUV fluxes approximately 1–2 orders of magnitude larger than that received by Earth today. Only $\delta$ Pav, 55 Cnc A, GL 183 and GJ 832 show HZ XUV environments comparable to the modern solar value, making them particularly interesting analogs for solar system–like irradiation conditions. We note that for stars with steep radial gradients in their high-energy emission, the flux can vary substantially across the HZ, such that planets at the inner edge may experience significantly enhanced atmospheric heating and escape compared to those near the outer edge. This implies that the range of XUV flux spanned by the HZ is comparable to the HZ width itself, and that planets within a single system may occupy markedly different photochemical and atmospheric evolution regimes despite all residing within the nominal habitable zone.

The FUV/NUV ratios of the solar-type stars are generally similar to solar values, although many are elevated by about 0.5 dex. In particular, $\delta$ Pav and 55 Cnc A show both XUV and FUV/NUV ratios similar to solar values, making them particularly compelling targets for solar system analog habitability studies. We note, however, that $\delta$ Pav is beginning to evolve off the main sequence, which may affect the long-term stability of its radiation environment and should be considered when evaluating it as a solar analog. 

Toward cooler stars, the FUV/NUV ratios systematically increase, reflecting relatively stronger FUV emission compared to NUV. This behavior is consistent with expectations that cooler stars have weaker photospheric NUV continua while maintaining chromospheric and transition-region FUV emission, resulting in radiation environments that may drive very different atmospheric photochemistry compared to Earth. Examining the stars with solar-like HZ XUV environments more closely, we find important differences in their photochemical environments. GJ 832 and GL 892 show FUV/NUV ratios roughly 1-2 orders of magnitude higher than solar, suggesting their HZ planets may experience substantially different atmospheric photochemistry despite having comparable XUV irradiation. These systems may therefore require more detailed photochemical modeling before being considered true solar analog radiation environments. At the same time, such elevated FUV/NUV environments may enhance the accumulation and detectability of key biosignature gases. Models show that reduced NUV flux relative to FUV prolongs the survival of biogenic species (e.g., CH$_4$, N$_2$O, CH$_3$Cl) and can enable up to an order-of-magnitude higher CH$_4$ abundances in oxygenated atmospheres around later-type K stars compared to solar-type systems, strengthening their potential observability \citep{Segura2005,Arney2019}. 

We also highlight the case of Lalande 21185, for which we constructed two SEDs using X-ray observations obtained in 2001 and 2004. These produce more than an order of magnitude difference in the inferred X-ray, EUV, and total XUV luminosities (Table \ref{tab:luminosities}; Figures \ref{fig:luminosity_scatter} and \ref{fig:HZ_fluxes}). This level of variability illustrates the importance of magnetic activity variability in M dwarfs and demonstrates that single-epoch observations may not fully characterize the long-term radiation environments relevant for habitability assessments.

Overall, these comparisons show that many stars in this sample exhibit higher high-energy radiation environments than the Sun, particularly at XUV wavelengths. However, it is important to emphasize that this represents a small and likely biased subset of potential HWO targets, comprising only about 6\% of the current high-priority target list. Furthermore, these stars are characterized using archival datasets, and in many cases were originally targeted for X-ray and UV observations specifically because they were already known or suspected to be magnetically active. As a result, this sample should not be interpreted as representative of the broader nearby solar-type star population.

The elevated EUV emission inferred for several stars will be examined further in Section \ref{subsec:euvcomparisons}, where we compare our model-derived EUV fluxes to empirical scaling relationships and to SEDs computed with alternate methods, and in Section \ref{subsec:origins}, where we investigate possible physical origins of the elevated XUV emission. Expanding high-energy observations to a more representative sample of HWO targets will be essential for determining whether these elevated radiation environments are typical or simply a consequence of selection effects.

These results also highlight the importance of obtaining new X-ray and UV observations of the remaining high-priority targets, including both single-epoch measurements to establish baseline activity levels and multi-epoch monitoring to quantify variability. The order-of-magnitude differences seen between the XUV emissions from the two Lalande 21185 SEDs illustrate how stellar variability can significantly alter inferred radiation environments, reinforcing the need for time-domain high-energy observations to robustly characterize the radiation conditions relevant for habitability assessments.

\begin{deluxetable*}{lccccccc}\label{tab:luminosities}
\tablecaption{Stellar High Energy and UV Luminosities}
\tablehead{
\colhead{Star} &
\colhead{$T_{\rm eff}$ (K)} &
\colhead{\shortstack{$\log L_{\rm X}$\\(1--100 $\AA$)}} &
\colhead{\shortstack{$\log L_{\rm EUV}$\\(100--912 $\AA$)}} &
\colhead{\shortstack{$\log L_{\rm FUV}$\\(1000--2000 $\AA$)}} &
\colhead{\shortstack{$\log L_{\rm NUV}$\\(2000--3000 $\AA$)}} &
\colhead{$\log L_{\rm bol}$} &
\colhead{$L_{\rm APEC}/L_{\rm EUV} (\%)$}
}
\startdata
$\iota$ Hor & 6298 & 29.43 & 29.16 & 30.35 & 32.07 & 33.81 & 67\\
$\chi$ Ori & 5906 & 29.49 & 29.45 & 29.91 & 31.74 & 33.58 & 67\\
GL 311 & 5962 & 29.71 & 29.58 & 30.13 & 31.79 & 33.59 & 75\\
GL 620A & 5818 & 29.11 & 28.73 & 29.65 & 31.60 & 33.55 & 47\\
$\kappa$ Cet & 5605 & 29.22 & 29.44 & 29.57 & 31.19 & 33.55 & 4\\
$\delta$ Pav & 5454 & 27.33 & 28.21 & 29.10 & 31.34 & 33.60 & 14\\
5 Cnc A & 5320 & 27.59 & 28.22 & 28.65 & 30.82 & 33.52 & 4\\
GL 892 & 4840 & 27.51 & 28.52 & 29.29 & 30.16 & 33.28 & 8\\
GL 183 & 4726 & 27.59 & 27.49 & 28.15 & 29.86 & 33.09 & 25\\
GJ 832 & 3657 & 26.98 & 26.74 & 27.54 & 27.83 & 32.19 & 27\\
Lalande 21185 & 3547 & $27.41-28.65$ & $27.54-28.78$ & 26.81 & 27.66 & 31.92 & $58-98$\\
Proxima Centauri & 2900 & 27.26 & 26.46 & 27.09 & 26.60 & 30.69 & 70\\
\hline
Sun & 5777 & $26.53-27.57$ & $27.68-28.22$ & $29.10-29.19$ & $31.60-31.60$ & 33.58 \\
\enddata
\tablenotetext{}{Note: The X-ray and EUV luminosity ranges reported for Lalande 21185 are derived from APEC models constructed from 2001 and 2004 observations. The final column gives the percentage of the integrated 100-912 \AA\ EUV flux contributed by the APEC coronal model, calculated as 100$\times$F$_{100-180}$/F$_{100-912}$. The remaining EUV flux is provided by the \texttt{PHOENIX} models.}
\end{deluxetable*}

\subsection{Comparison to Published SEDs and EUV Predictions}\label{subsec:euvcomparisons}

Because EUV wavelengths are currently inaccessible to direct observation due to strong attenuation by the interstellar medium and the lack of operational instrumentation covering this spectral region, indirect reconstruction methods are required to estimate stellar EUV emission. Several groups have developed approaches to estimate EUV fluxes, including semi-empirical SED construction and scaling relationships based on observable high-energy diagnostics such as X-ray and UV emission lines. To assess the reliability of our reconstructed EUV fluxes, we compare our SEDs to previously published stellar SEDs where available and evaluate our EUV predictions against a variety of empirical scaling relationships. These comparisons provide an important consistency check on our methodology and help place our EUV estimates in the broader context of existing reconstruction techniques.

\subsubsection{Comparison to Literature SEDs}

Published panchromatic SEDs suitable for direct comparison exist for three stars in our sample: $\kappa^1$ Ceti and the M dwarfs GJ 832 and Proxima Centauri.

For GJ 832, our reconstructed SED yields an EUV luminosity of $\log L_{\rm EUV} = 26.74$ erg s$^{-1}$, which is modestly lower than previous estimates. \cite{Peacock2019} derived a value of $\log L_{\rm EUV} = 27.31$ erg s$^{-1}$ using an earlier version (V16.09) of the \texttt{PHOENIX} code guided by the same HST UV observations but without the inclusion of an APEC coronal model. The difference in EUV luminosity is primarily attributable to updates in the PHOENIX models. The present analysis uses PHOENIX V20.01 and includes an expanded treatment of non-LTE species and more complete atomic line lists. Specifically, the number of atoms and ions treated in non-LTE increased from 73 to 78, the number of levels from 15,355 to 19,879, and the number of emission lines from 233,871 to 377,034. Many of the additional lines that are now treated in non-LTE and were previously treated in LTE occur at EUV wavelengths and were overpredicted in the earlier model. The semi-empirical Solar-Stellar Radiation Physical Model (SSRPM) approach of \cite{Fontenla2013} produces a similar value of $\log L_{\rm EUV} = 27.26$ erg s$^{-1}$, while the MUSCLES survey SED from \cite{France2016}, which estimates EUV flux using the \cite{Linsky2014} $F_{\rm EUV}$–$F_{\rm Ly\alpha}$ scaling relationship, finds $\log L_{\rm EUV} = 27.39$ erg s$^{-1}$. 

For Proxima Centauri, our SED produces an EUV luminosity of $\log L_{\rm EUV} = 26.51$ erg s$^{-1}$, compared to $\log L_{\rm EUV} = 27.05$ erg s$^{-1}$ from the MUSCLES SED of \cite{France2016}, which again relied on the \cite{Linsky2014} Ly$\alpha$-based EUV scaling relationship. For $\kappa^1$ Cet, our reconstructed EUV luminosity of $\log L_{\rm EUV} = 29.43$ erg s$^{-1}$ is slightly higher than the $\log L_{\rm EUV} = 29.05$ erg s$^{-1}$ derived by \cite{Wilson2026} using a differential emission measure (DEM) model.

Overall, these comparisons show that our computed EUV luminosities can differ from previously published values by a few tenths of a dex, with the direction and magnitude of the difference depending on the reconstruction method and available observational constraints. The generally good agreement across independent approaches supports the reliability of our SED reconstructions while highlighting the systematic uncertainties inherent to estimating stellar EUV emission in the absence of direct observations.

\subsubsection{Comparison to EUV Scaling Relationships}\label{subsub:scalings}

In addition to literature SED comparisons, we also compare our reconstructed EUV luminosities to predictions from a variety of empirical scaling relationships. These relations span a range of methodologies, including correlations between X-ray and EUV emission based on stellar coronal modeling, relations derived from solar irradiance variability extended to stellar regimes, and proxies based on FUV transition region emission lines\footnote{Note: We do not compare to EUV estimates derived from the commonly used \citealt{Linsky2014} Ly$\alpha$ scaling relation because our models adopt the self-consistent \texttt{PHOENIX} Ly$\alpha$ profiles rather than reconstructed intrinsic profiles. Differences could therefore arise from the Ly$\alpha$ modeling, the EUV reconstruction, or the scaling relation itself, making interpretation ambiguous.}. We provide descriptions of each method in Appendix~\ref{appendix:euv_scalings}. Together these approaches provide independent estimates of EUV emission using observables that are more readily accessible than direct EUV measurements.

The scaling relations estimate EUV emission across a range of wavelength bands, spanning roughly 80–920~\AA, allowing us to test whether discrepancies arise in particular spectral regions or from specific components of our SED construction (e.g., the APEC coronal models dominating shortward of 180~\AA\ versus the \texttt{PHOENIX} chromosphere and transition region models at longer EUV wavelengths). We present these comparisons in Figure~\ref{fig:euv_comps} and list the predicted and SED-derived values in Table~\ref{tab:euvcomps} in the Appendix.

Overall, we find good general agreement between our reconstructed EUV luminosities and the scaling relationships, particularly for the FGK stars. For these stars, the mean ratios between scaling predictions and SED-derived EUV luminosities are typically within a few percent of unity (Table~\ref{tab:euvcomps}), and nearly all values fall within the intrinsic 0.3-0.5 dex scatter of the empirical relations. This level of agreement suggests that our semi-empirical SED construction produces EUV fluxes consistent with established activity scalings. GL 892 also appears less anomalous in this context, as its reconstructed EUV luminosity agrees with X-ray–based scaling predictions, suggesting its relatively high EUV output compared to solar values (Figure \ref{fig:luminosity_scatter}) may not be solely an artifact of the limited UV constraints used in the SED reconstruction.

For the three M dwarfs (GJ~832, Proxima Centauri, and Lalande 21185), our SED-derived EUV luminosities tend to fall systematically below predictions from the X-ray-based scaling relations (e.g., \citealt{Johnstone2021} and \citealt{Chadney2015}), while remaining broadly consistent with the FUV emission-line based estimates of \citealt{France2018} for the two stars with available \ion{N}{5} and \ion{Si}{4} measurements (GJ 832 and Proxima Centauri). This behavior is particularly interesting because the \citealt{France2018} relations predict EUV emission over 90–360~\AA, a region in our SEDs that includes contributions from both the APEC coronal models and the FUV-constrained \texttt{PHOENIX} atmospheric structure. The relative contribution of these components varies substantially among the M dwarfs: APEC accounts for approximately 54\% of the 90--360~\AA\ flux for GJ~832, but more than 97\% for both Proxima Centauri and Lalande 21185. Thus, the agreement with the \citealt{France2018} relation for GJ~832 reflects contributions from both the coronal and upper-atmospheric components, whereas the agreement for Proxima Centauri (note: Lalande 21185 does not have observed \ion{N}{5} or \ion{Si}{4} line fluxes that allowed for comparison with these scaling relationships) is predominantly determined by the coronal emission model. The fact that Proxima Centauri remains consistent with the FUV-based relation despite its 90--360~\AA\ emission being almost entirely determined by the APEC component suggests that the lower EUV luminosities derived here are not simply a consequence of the FUV-constrained \texttt{PHOENIX} models. Instead, they reflect the combined X-ray-to-EUV spectral structure of the reconstructed SEDs and may indicate that X-ray-based scaling relations overpredict the EUV emission for these active M dwarfs.

The two Lalande 21185 SEDs provide an additional test of the effects of stellar variability. Across nearly all scaling relations, at least one of the two SED realizations falls within the expected RMS scatter of the relations (Table~\ref{tab:euvcomps}, Appendix). In the comparisons that isolate the \texttt{PHOENIX} contribution to the EUV emission from the X-ray-derived component (Figure \ref{fig:euv_comps}, top and middle right), the SED based on the lower 2001 X-ray flux epoch generally shows better agreement with the scaling relations. This suggests that the lower X-ray state may provide a more representative estimate of the stellar UV/EUV emission than the higher 2004 X-ray state. However, the X-ray and UV observations are not contemporaneous: the two X-ray epochs are from 2001 and 2004, whereas the FUV observations were obtained in 2019--2020 and the NUV observations in 2022. Given the suggested $\sim$7 yr activity cycle for Lalande 21185, these observations may sample substantially different phases of the stellar activity cycle, preventing us from directly associating either X-ray epoch with the observed UV flux. The preference for the lower X-ray SED should therefore be interpreted as suggestive rather than as evidence that the 2001 or 2004 X-ray state more closely represents the UV-observed activity level. Nevertheless, this comparison demonstrates that the observed level of X-ray variability can produce substantial differences in the reconstructed EUV spectrum and reinforces the importance of accounting for activity variability when applying scaling relations to M dwarfs, where single-epoch measurements may not be representative of typical high-energy output.

Examining the scaling relations by wavelength coverage provides additional insight. For the broad EUV bands (e.g., the 100–920~\AA\ and 100–504~\AA\ relations from \citealt{SanzForcada2025} and the 62–912~\AA\ relation from \citealt{King2018}), which probe EUV emission arising from both the coronal and transition region portions of our SEDs, we find consistently good agreement, although our SEDs tend to predict slightly lower EUV luminosities than the \citealt{SanzForcada2025} relations. The agreement with the \citealt{King2018} relation is particularly strong, with nearly all stars falling within the expected scatter envelope (Figure~\ref{fig:euv_comps}).

The intermediate 200–504~\AA\ comparisons between the \citealt{SanzForcada2025} and \citealt{Poppenhaeger2022} relations show that our FGK stars tend to fall between these two relations, while the M dwarfs fall below both predictions. This likely reflects differences in how these relations were constructed, including the use of coronal emission measure modeling versus direct EUV measurements, as well as known challenges associated with EUV observations in this wavelength region.

The short-wavelength EUV comparisons (80–360~\AA), including the \citealt{Johnstone2021}, \citealt{Chadney2015}, and \citealt{France2018} relations, reveal the clearest systematic behavior. Our SED values tend to fall slightly below the X-ray-based predictions from \citealt{Johnstone2021} and \citealt{Chadney2015}, while falling closer to or slightly above the FUV-based predictions from \citealt{France2018}. Because this wavelength range includes the 80–180~\AA\ portion modeled with APEC, these comparisons are sensitive to the adopted coronal emission model. Indeed, the APEC component accounts for the majority of the 80–360~\AA\ flux for several stars, including $\iota$ Hor, $\chi^1$ Ori, GL~311, GL~620-1A, Lalande 21185, and Proxima Centauri, with contributions ranging from approximately 80\% to $>97\%$. Notably, the F and early G stars lie above the RMS scatter envelope in the France \ion{N}{5} relation while remaining consistent with the other scalings, suggesting that this particular proxy may be less reliable for hotter stars.

Finally, the \citealt{Johnstone2021} 360–920~\AA\ relation, which corresponds directly to the \texttt{PHOENIX}-dominated portion of our EUV spectra, shows particularly strong agreement with our computed values. This provides additional confidence that the chromospheric and transition-region structure in our \texttt{PHOENIX} models produces realistic EUV emission levels.

Taken together, these comparisons indicate that our EUV luminosities are broadly consistent with existing empirical scaling relationships, with typical differences well within the intrinsic scatter of the relations. The most notable deviations occur for the M dwarfs, where variability and the limitations of X-ray-based scalings likely play an important role. Overall, this agreement provides an important validation that our semi-empirical SED methodology produces EUV fluxes consistent with independent reconstruction techniques.

\begin{figure*}
    \centering
    \includegraphics[width=0.95\linewidth]{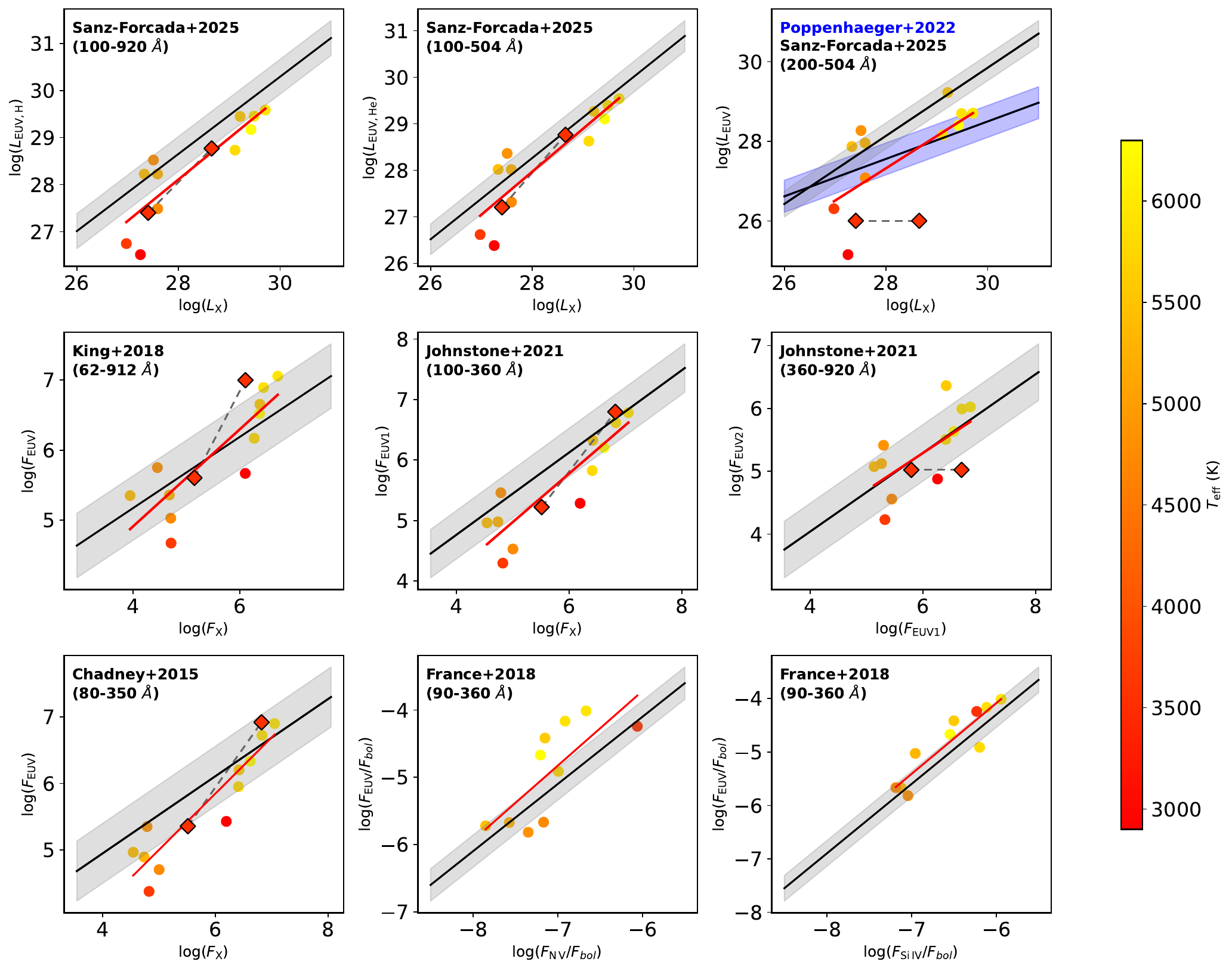}
    \caption{Comparison of EUV fluxes derived from the panchromatic SEDs to predictions from empirical scaling relationships. Panel labels indicate the wavelength ranges used for each scaling relationship. Each panel shows integrated EUV fluxes or luminosities from this work (points, color–coded by effective temperature) compared to the published relations, shown as black lines, with gray shaded regions indicating the approximate RMS scatter reported for each relationship. Diamond symbols connected by a dashed line are the values from the two Lalande 21185 SEDs with APEC models constructed from 2001 and 2004 observations. Red lines show linear fits to the SED-derived values. In the top right panel, both the \citet{SanzForcada2025} relation (black) and the earlier \citet{Poppenhaeger2022} relation (blue) are shown, as both predict EUV emission over the same wavelength range. A detailed numerical comparison between the EUV fluxes derived from our SEDs and those predicted by the scaling relations is provided in Appendix Table \ref{tab:euvcomps}.}
    \label{fig:euv_comps}
\end{figure*}

\subsection{Origin of the Elevated XUV Emission}\label{subsec:origins}

\begin{figure*}
    \centering
    \includegraphics[width=0.95\linewidth]{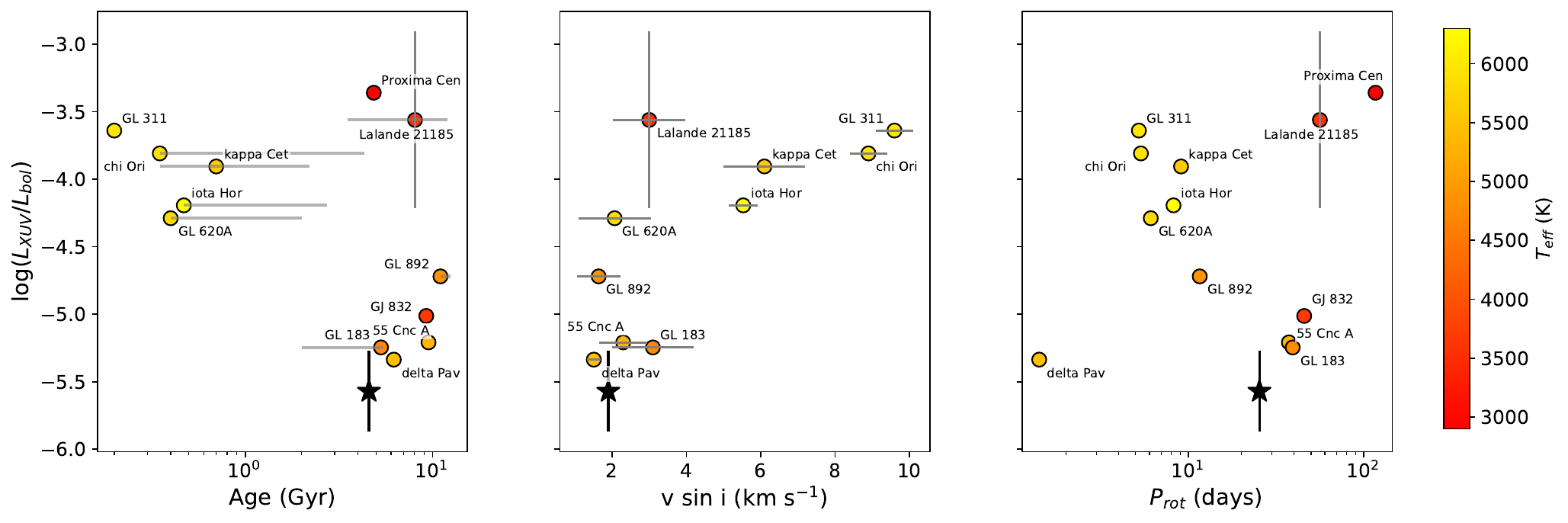}
    \caption{Stellar high-energy activity, expressed as $\log(L_{\rm XUV}/L_{\rm bol})$, plotted against stellar age (left), projected rotational velocity (center), and rotation period (right) (values given in Table \ref{tab:activity}).  Symbols are colored by effective temperature. Vertical error bars show the range of XUV luminosities derived from multiple epochs for Lalande 21185 and the Sun, reflecting activity variability, while horizontal error bars denote uncertainties in $v\sin i$. The Sun is plotted as a black star for comparison.}
    \label{fig:activity_comps}
\end{figure*}

Many stars in our sample exhibit fractional XUV luminosities exceeding that of the Sun (Figure \ref{fig:activity_comps}). This likely reflects a combination of sample selection effects, differences in stellar rotation and age, and intrinsic magnetic variability rather than indicating that the Sun is unusually inactive.

An important contributor is sample bias. Our targets do not represent a volume-limited sample of solar analogs, but instead consist largely of stars previously observed at high energies, which naturally favors magnetically active stars. This is illustrated by comparison to the broader TSS25 target list, where only 51 of 164 high-priority targets currently have X-ray observations. Of these, 22 have only faint detections and 8 are unresolved binaries, leaving just 19 stars with robust X-ray detections. The comparable number of faint detections to those with robust detections (that our sample is pulled from) suggests that many stars with more solar-like activity levels remain poorly characterized at high energies, and that our sample likely represents the more active subset of the population.

Stellar rotation and age also play an important role. As shown in Figure \ref{fig:activity_comps}, stars with shorter rotation periods and larger projected rotational velocities generally show higher $\log(L_{\rm XUV}/L_{\rm bol})$, consistent with the well-established rotation–activity relationship. Several solar-type stars in our sample are substantially younger than the Sun and rotate significantly faster, naturally leading to stronger magnetic dynamos and enhanced high-energy emission. In contrast, the oldest and most slowly rotating stars in the sample, including $\delta$ Pav, 55 Cnc A, and GL 892, exhibit activity levels comparable to or somewhat higher than the present-day solar range. Although the Sun's maximum activity level overlaps with that of comparable stars, its average level is lower, suggesting that the Sun may be relatively quiet compared with its stellar peers.

Magnetic variability may also contribute to some of the observed differences. The X-ray variability analysis done in \citealt{Binder2024} shows that while most solar-type stars in our sample exhibit modest variability (typically $\lesssim0.5$ dex), several late-type stars show much larger variations. In particular, GJ 832 and Proxima Centauri vary by about an order of magnitude in X-ray luminosity. This is especially relevant for Proxima Centauri, where the X-ray data represent an average over observations spanning nearly two decades, while the UV observations were obtained near the beginning of this interval. Similar temporal offsets between UV and X-ray observations may introduce additional scatter in the inferred XUV luminosities for other stars as well.

Finally, elevated XUV environments relative to the Sun appear to be common among well-characterized exoplanet host stars. For example, \citet{Wilson2026} found that most FGKM transiting exoplanet host stars observed in JWST Cycle 1 exhibit XUV fluxes at their habitable zones that exceed solar values by 1–2 orders of magnitude, with substantial scatter even among stars of similar effective temperature. Together, these results suggest that the elevated XUV emission seen in our sample reflects both the underlying physics of stellar magnetic activity and observational biases toward stars with detectable high-energy emission.

\subsection{Implications for Habitable Zone Planets}

The elevated XUV fluxes of the stars in our sample raise the question of whether planets located in their HZs are subject to atmospheric escape rates large enough to negatively impact their long-term habitability. At HZ distances, the stellar spectral energy distributions presented here typically produce XUV fluxes that are 1–2 orders of magnitude higher than the present-day solar value at Earth. However, many of the solar-type stars in our sample are younger than the Sun, and their elevated high-energy emission is therefore consistent with the expected age–activity–rotation relation. Results from the Sun in Time program further support this interpretation, showing that the young Sun likely produced X-ray and EUV emission 100–1000 times stronger than today, while FUV and NUV emission may have been enhanced by factors of roughly 20–60 and 10–20, respectively \citep{Ribas2005}. These levels are comparable to the range of XUV environments found in our sample and to those experienced by the early Earth when the Sun was significantly more magnetically active. This comparison suggests that enhanced high-energy irradiation alone does not necessarily imply uninhabitable conditions.


Several stars in this sample host known planetary systems, providing concrete examples of how these radiation environments may affect real HZ planets. Five of the twelve stars have confirmed planets, including systems dominated by gas giants (e.g., $\iota$ Hor and GJ 832), compact multi-planet super-Earth systems (e.g., GL 892), and the well-studied Proxima Centauri system. While most known planets in these systems are either short-period planets unlikely to be habitable or gas giants, some provide useful context for interpreting the radiation environments derived here. For example, the gas giant 55 Cnc f lies near the outer edge of the habitable zone and could potentially host habitable moons; notably, we find that the XUV flux at the HZ of 55 Cnc is comparable to that received by the Earth from the modern Sun, suggesting a relatively benign high-energy radiation environment compared to most stars in this sample. In contrast, Proxima Centauri b represents one of the best known terrestrial-mass HZ planets around a nearby star. Our SEDs indicate that Proxima Centauri b receives roughly two orders of magnitude more XUV flux than the Earth does today, and this planet is also subject to frequent stellar flaring known to occur on Proxima Centauri, both of which may contribute to enhanced atmospheric escape and photochemical evolution. However, as discussed above, such comparisons must also consider the time evolution of stellar activity and the potential for atmospheric retention or replenishment. 

\begin{figure*}
    \centering
    \includegraphics[width=0.75\linewidth]{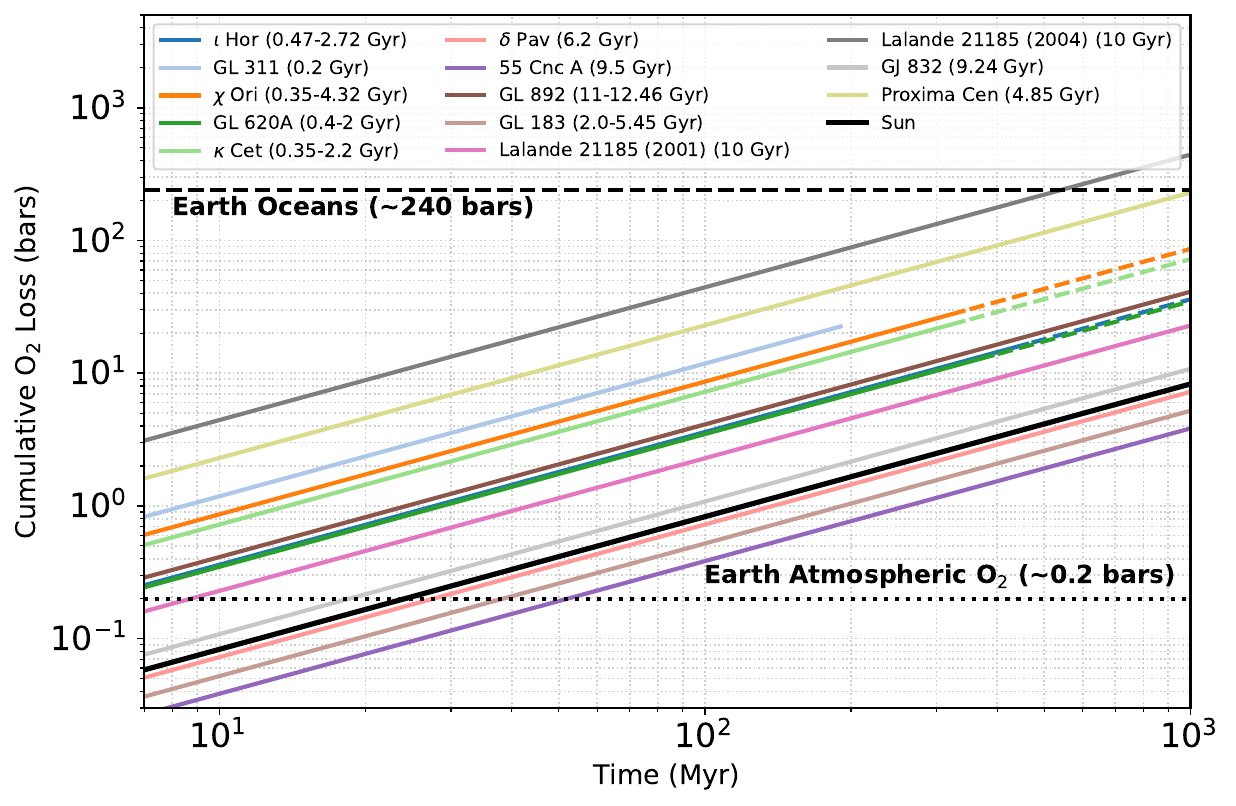}
    \caption{Cumulative oxygen loss for planets at the habitable zone of each sample star over 1 Gyr, assuming constant stellar XUV flux at the HZ and no replenishment of oxygen. The curves are calculated using the energy-limited O$_2$ loss prescription from \citealt{Airapetian2017}. Colored lines show individual stars, with solid segments representing the earliest ages (for stars with age ranges) and dashed segments extending beyond the minimum age. The black solid line shows the solar reference. Horizontal lines indicate the present-day Earth atmospheric O$_2$ content ($\sim$0.2 bars; dotted) and the total mass of Earth’s oceans ($\sim$240 bars; dashed), providing reference levels for planetary habitability.}
    \label{fig:o2loss}
\end{figure*}

To more broadly quantify the potential impact of these elevated fluxes on the HZs of the sample stars, we estimate cumulative oxygen losses using the scaling relation from \citet{Airapetian2017}, assuming constant XUV irradiation at the HZ over a 1 Gyr interval (Figure \ref{fig:o2loss}). For context, we show reference levels corresponding to the oxygen content of Earth's present atmosphere ($\sim$0.2 bars) and the oxygen equivalent of Earth's oceans ($\sim$240 bars). Under these assumptions, most FGK stars in our sample would drive the loss of only a few to several tens of bars of oxygen over a Gyr, while the solar case lies near the lower end of this range. In contrast, the highest XUV flux M dwarfs approach or exceed the equivalent of an Earth ocean, suggesting that total volatile loss may be more significant for planets orbiting the most active low-mass stars.

These estimates should be regarded as conservative upper limits. The adopted scaling assumes constant XUV flux, whereas stellar high-energy emission declines with age following rotational spin-down. In reality, most atmospheric escape would occur early in a system's evolution, producing curved rather than linear cumulative loss histories. Furthermore, these calculations neglect atmospheric replenishment processes such as volcanic outgassing, impact delivery, and interior–atmosphere exchange, all of which could substantially mitigate net volatile loss. The results shown here therefore represent pessimistic, worst-case scenarios rather than expected evolutionary outcomes.

Beyond atmospheric escape, the elevated XUV environments characterized here also bear on whether HZ planets may evolve toward Venus-like climate states. The runaway greenhouse represents an alternative pathway in which enhanced stellar irradiation triggers the complete evaporation of surface water, leading to a steam atmosphere, rapid hydrogen loss, and eventual desiccation \citep{ingersoll1969c,kasting1988c,hamano2013,goldblatt2013,turbet2021}. This outcome is particularly relevant for planets near the inner edge of the HZ, where the boundary between Venus and Earth evolutionary trajectories is defined by the Venus Zone \citep{kane2014e,ostberg2023a}. For the younger, more active stars in our sample, elevated XUV fluxes during the pre-main-sequence and early main-sequence phases may have driven early ocean loss on inner HZ planets before surface conditions could stabilize; an outcome that 3D climate models suggest may have previously occurred on Venus \citep{way2020,turbet2021}. For planets orbiting M dwarfs, prolonged pre-main-sequence luminosity further compounds this risk by extending the period during which inner HZ planets receive super-critical insolation, potentially producing abiotic oxygen buildup as a false positive for biological activity \citep{luger2015b}. As emphasized by \citet{kane2024b}, Venus provides a critical anchor point for interpreting the habitability of terrestrial exoplanets, and the synergies between Venus science and exoplanet characterization \citep{kane2019d,way2023a}. Thus, it is important to consider exoVenus outcomes alongside temperate scenarios when using the SEDs presented here to model HZ planetary atmospheres for HWO and ELT target prioritization \citep{kane2026a}.

Taken together, these results suggest that although planets in the HZs of these stars are exposed to systematically higher XUV radiation fields than modern Earth, catastrophic and complete volatile loss is not an inevitable consequence. For most FGK stars in this sample, the estimated losses remain well below an Earth ocean equivalent even under pessimistic assumptions. The greatest potential risk appears to be confined to planets orbiting the most XUV-active M dwarfs, where sustained high-energy irradiation could significantly alter combined atmospheric and surface inventories. However, even the modestly elevated XUV radiation environments of most FGK stars could influence the viability of truly Earth-analog N$_{2}$-O$_{2}$ dominated atmospheres with low CO$_{2}$ \citep{Johnstone2021b, Sherf2024}. Overall, these findings indicate that elevated XUV environments alone are unlikely to preclude habitability for most HZ planets in this sample, but they may represent an important evolutionary pressure that must be considered alongside planetary mass, atmospheric composition, and geophysical replenishment processes.

\section{Conclusions}\label{sec:summary}

We have constructed panchromatic SEDs spanning the X-ray through radio for 12 high-priority target stars identified for future direct imaging efforts with HWO and the ELTs. These stars comprise the 6\% of targets in the TSS25 sample for which sufficiently complete archival X-ray and ultraviolet observations currently exist.

We find that many stars in this sample exhibit systematically elevated high-energy radiation environments relative to the modern Sun, with XUV fluxes at their HZs frequently exceeding solar values by one to two orders of magnitude. Comparisons with independent empirical EUV scaling relationships demonstrate that these elevated fluxes are not artifacts of the \texttt{PHOENIX} atmosphere models used to construct the SEDs. The agreement between model-derived EUV fluxes and predictions based on both X-ray and UV diagnostics provides confidence in the robustness of the inferred high-energy radiation environments, although the M-dwarf models tend to produce slightly lower EUV fluxes than empirical estimates.

The origin of this enhanced activity likely reflects a combination of factors rather than a single physical cause. Because the sample is limited to stars with existing high-energy observations, it is naturally biased toward magnetically active targets. In addition, variations in stellar age, rotation, and magnetic cycle phase likely contribute to the observed spread in activity levels. Intrinsic variability may also play a major role. The order-of-magnitude differences observed between the XUV emissions of two Lalande 21185 SED realizations demonstrate how strongly time variability can affect the inferred radiation environment of potentially habitable planets.

Comparison with the early solar radiation environment suggests that these elevated XUV levels do not necessarily preclude habitability. Many of the inferred HZ fluxes are comparable to those experienced by the early Earth, when solar high-energy emission was substantially stronger than today. Estimating the oxygen loss from their HZ planets, we find that most FGK stars in this sample would drive the loss of at most a few to several tens of bars of oxygen over a Gyr assuming constant irradiation and no atmospheric replenishment, remaining well below an Earth ocean equivalent even under these pessimistic assumptions. Only the most XUV-active M dwarfs approach losses comparable to an Earth ocean over similar timescales. These results suggest that while enhanced XUV irradiation may accelerate atmospheric evolution, it is unlikely to represent a prohibitive barrier to habitability for most HZ planets in this sample when considered in the broader context of planetary evolution and volatile replenishment processes.

These results underscore a major limitation in our current understanding of the high-energy environments of future direct imaging targets: only a small fraction of the highest-priority stars currently have adequate X-ray and UV characterization. The present sample represents just 9 of the 164 stars in the TSS25 catalog, highlighting the need for substantial expansion of high-energy observations. Both single-epoch observations to establish baseline activity levels and multi-epoch monitoring to quantify variability will be essential for developing realistic radiation environment estimates.

More broadly, identifying the most promising targets for habitability studies will require improved constraints on stellar age, rotation, and magnetic activity evolution. In particular, additional observations are needed to identify older, slowly rotating solar-type stars with lower X-ray and FUV emission that may provide more benign radiation environments for habitable zone planets. Expanding the high-energy observational foundation for the full HWO target list will therefore be critical not only for interpreting future exoplanet atmosphere measurements but also for prioritizing the stars most likely to host planets capable of retaining habitable environments.

\section{Acknowledgments}
This work is supported by NASA Exoplanets Research Program (XRP) award \#80NSSC23K0039 (PI Turnbull). E.S. and S.P. acknowledge support from the CHAMPs (Consortium on Habitability and Atmospheres of M-dwarf Planets) team, supported by the National Aeronautics and Space Administration (NASA) under grant nos. 80NSSC21K0905 and 80NSSC23K1399, issued through the Interdisciplinary Consortia for Astrobiology Research (ICAR) program. S.P. also acknowledges support from NASA under award number 80GSFC24M0006. This work uses observations obtained with the NASA/ESA Hubble Space Telescope, retrieved from the Mikulski Archive for Space Telescopes (MAST) at the Space Telescope Science Institute (STScI). STScI is operated by the Association of Universities for Research in Astronomy, Inc. under NASA contract NAS 5-26555. The results presented in this document also rely on data measured from the Thermosphere Ionosphere Mesosphere Energetics and Dynamics (TIMED) Solar EUV Experiment (SEE). These data are available from the TIMED SEE website at https://lasp.colorado.edu/see/data/. These data were accessed via the LASP Interactive Solar Irradiance Datacenter (LISIRD) (https://lasp.colorado.edu/lisird/).

\appendix

\section{EUV Scaling Relationships}\label{appendix:euv_scalings}

To place our reconstructed EUV luminosities in the context of existing estimation techniques, we compare our SED-derived values to predictions from several commonly used EUV scaling relationships drawn from the literature. These relations use a variety of observational proxies, including X-ray luminosity, X-ray surface flux, solar irradiance trends, and FUV transition-region emission lines, and typically show intrinsic dispersions of $\sim$0.2–0.5 dex. Table~\ref{tab:euvcomps} summarizes the EUV luminosities derived from our SEDs alongside the values predicted by each scaling relation. Below we briefly describe the methodology and applicability of each relation to provide context for the comparison. 

\textbf{Sanz-Forcada+2025}\\
\citet{SanzForcada2025} derived EUV scaling relations using synthetic XUV spectra generated from coronal emission-measure models. The sample consists of $\sim$75 F–M stars with X-ray observations primarily from \textit{XMM-Newton} and \textit{Chandra}, supplemented by UV emission lines from \textit{HST}, \textit{FUSE}, and archival \textit{EUVE} data. Differential emission measure distributions were constructed from observed coronal and transition-region lines and used to compute synthetic spectra across the XUV range. EUV luminosities were then integrated over 100–504~\AA\ (helium-ionizing) and 100–920~\AA\ (hydrogen-ionizing) bands and related to $L_X$ via log-linear regressions, with typical scatter of $\sim0.3$–$0.4$ dex. The authors also derived a revised relation between $L_X$ and the 200–504~\AA\ band to revisit the relation of \citet{Poppenhaeger2022}, which they argue is unreliable because the EUV fluxes were derived directly from \textit{EUVE} spectra in a wavelength region strongly affected by interstellar absorption, instrumental limitations, and line-identification issues. However, differences between the relations may also arise from the use of coronal emission-measure modeling rather than direct EUV observations.

\textbf{Poppenhaeger+2022}\\
\citet{Poppenhaeger2022} derived empirical relations between stellar X-ray and EUV luminosities using nearby late-type stars with both X-ray and EUV observations. X-ray luminosities were compiled from \textit{XMM-Newton}, \textit{Chandra}, and archival \textit{ROSAT} measurements, while EUV luminosities were obtained from archival \textit{EUVE} spectra and integrated over the 200–504~\AA\ wavelength range. Power-law relations were fit between $\log L_{\rm EUV}$ and $\log L_X$, allowing EUV emission to be estimated from commonly available X-ray measurements. The typical dispersion of the relation is $\sim0.3$–$0.4$ dex in $\log L_{\rm EUV}$.

\textbf{King+2018}\\
\citet{King2018} derived X-ray–EUV scaling relations based on solar irradiance measurements over the solar activity cycle. Solar spectral irradiance from the TIMED/SEE mission was used to measure simultaneous X-ray and EUV fluxes, which were then expressed in terms of surface flux and extrapolated to stellar activity levels. EUV fluxes were integrated over the 100–912~\AA\ band and related to X-ray surface flux through a power-law relation between $F_{\rm EUV}/F_X$ and $F_X$. When applied to stellar samples, the relation exhibits a typical scatter of $\sim0.4$–$0.5$ dex.

\textbf{Johnstone+2021}\\
\citet{Johnstone2021} derived relations between stellar X-ray surface flux and EUV surface flux using coronal modeling of late-type stars. X-ray measurements were compiled from \textit{XMM-Newton}, \textit{Chandra}, and \textit{ROSAT} observations spanning a range of stellar activity levels. Synthetic XUV spectra were generated from coronal emission measure distributions constrained by the X-ray data, and EUV surface fluxes were integrated over two wavelength intervals: 100-360~\AA\ (EUV1) and 360–920~\AA\ (EUV2). Power-law relations were then fit between $F_X$ and $F_{\rm EUV}$. The EUV2 relation is derived primarily from solar data and is therefore considered less reliable for stars. The typical scatter in the relations is $\sim0.4$ dex.

\textbf{Chadney+2015}\\
\citet{Chadney2015} developed a relation between stellar X-ray surface flux and EUV surface flux using solar observations combined with a small sample of active stars. Solar XUV spectra from the TIMED/SEE mission were used to determine how the ratio $F_{\rm EUV}/F_X$ varies with X-ray surface flux across the solar activity cycle. This trend was then extended to stellar activity levels using X-ray measurements of nearby late-type stars obtained with \textit{XMM-Newton} and \textit{ROSAT}. EUV fluxes were defined over the 100–912~\AA\ band, and the resulting relation predicts EUV flux from X-ray surface flux with a typical uncertainty of $\sim0.35$–$0.4$ dex.

\textbf{France+2018}\\
\citet{France2018} derived empirical scaling relations between far-ultraviolet (FUV) transition-region emission lines and stellar EUV luminosity. The proxies used were the \ion{N}{5} $\lambda\lambda1238,1242$ and \ion{Si}{4} $\lambda\lambda1393,1402$ doublets measured with \textit{HST}/COS and \textit{HST}/STIS for nearby F–M stars with existing EUV observations. EUV luminosities in the 90–360~\AA\ band were obtained from archival \textit{EUVE} spectra and calibrated against the FUV line luminosities using relations between fractional luminosities (e.g., $\log(L_{\rm EUV}/L_{\rm bol})$ vs.\ $\log(L_{\rm line}/L_{\rm bol})$). The resulting relations exhibit an intrinsic scatter of $\sim0.23$–$0.25$ dex.

\begin{deluxetable*}{lccccccccccc}\label{tab:euvcomps}
\tabletypesize{\scriptsize}
\setlength{\tabcolsep}{4pt}
\renewcommand{\arraystretch}{1.35}
\tablehead{Star & \shortstack{Sanz-Forcada\\$\log L_{\rm EUV,H}$\\(100--920\,\AA)} & \shortstack{Sanz-Forcada\\$\log L_{\rm EUV,He}$\\(100--504\,\AA)} & \shortstack{Sanz-Forcada\\$\log L_{\rm EUV}$\\(200--504\,\AA)} & \shortstack{Poppenhaeger\\$\log L_{\rm EUV}$\\(200--504\,\AA)} & \shortstack{King\\$\log F_{\rm EUV}$\\(62--912\,\AA)} & \shortstack{Johnstone\\$\log(F_{\rm EUV1})$\\(100--360\,\AA)} & \shortstack{Johnstone\\$\log(F_{\rm EUV2})$\\(360--920\,\AA)} & \shortstack{Chadney\\$\log F_{\rm EUV}$\\(80--350\,\AA)} & \shortstack{France (N V)\\$\log(F_{\rm EUV}/F_{\rm bol})$\\(90--360\,\AA)} & \shortstack{France (Si IV)\\$\log(F_{\rm EUV}/F_{\rm bol})$\\(90--360\,\AA)} & \shortstack{Mean\\ratio}}
\startdata
iotaHor & \shortstack[c]{\rule{0pt}{10pt}29.79\\29.16\\1.02} & \shortstack[c]{\rule{0pt}{10pt}29.48\\29.09\\1.01} & \shortstack[c]{\rule{0pt}{10pt}29.36\\28.40\\1.03} & \shortstack[c]{\rule{0pt}{10pt}28.23\\28.40\\0.99} & \shortstack[c]{\rule{0pt}{10pt}6.38\\6.52\\0.98} & \shortstack[c]{\rule{0pt}{10pt}6.55\\6.20\\1.06} & \shortstack[c]{\rule{0pt}{10pt}5.68\\5.63\\1.01} & \shortstack[c]{\rule{0pt}{10pt}6.47\\6.33\\1.02} & \shortstack[c]{\rule{0pt}{10pt}5.65\\6.28\\0.90} & \shortstack[c]{\rule{0pt}{10pt}5.94\\6.28\\0.95} & 1.00\\
chiOri & \shortstack[c]{\rule{0pt}{10pt}29.84\\29.45\\1.01} & \shortstack[c]{\rule{0pt}{10pt}29.53\\29.39\\1.00} & \shortstack[c]{\rule{0pt}{10pt}29.41\\28.70\\1.02} & \shortstack[c]{\rule{0pt}{10pt}28.26\\28.70\\0.98} & \shortstack[c]{\rule{0pt}{10pt}6.41\\6.89\\0.93} & \shortstack[c]{\rule{0pt}{10pt}6.69\\6.61\\1.01} & \shortstack[c]{\rule{0pt}{10pt}5.81\\5.99\\0.97} & \shortstack[c]{\rule{0pt}{10pt}6.59\\6.72\\0.98} & \shortstack[c]{\rule{0pt}{10pt}5.83\\6.67\\0.87} & \shortstack[c]{\rule{0pt}{10pt}6.39\\6.67\\0.96} & 0.98\\
GL311 & \shortstack[c]{\rule{0pt}{10pt}30.02\\29.58\\1.01} & \shortstack[c]{\rule{0pt}{10pt}29.72\\29.53\\1.01} & \shortstack[c]{\rule{0pt}{10pt}29.60\\28.71\\1.03} & \shortstack[c]{\rule{0pt}{10pt}28.36\\28.71\\0.99} & \shortstack[c]{\rule{0pt}{10pt}6.55\\7.05\\0.93} & \shortstack[c]{\rule{0pt}{10pt}6.84\\6.78\\1.01} & \shortstack[c]{\rule{0pt}{10pt}5.95\\6.02\\0.99} & \shortstack[c]{\rule{0pt}{10pt}6.72\\6.89\\0.97} & \shortstack[c]{\rule{0pt}{10pt}6.09\\6.84\\0.89} & \shortstack[c]{\rule{0pt}{10pt}6.63\\6.84\\0.97} & 0.98\\
GL620A & \shortstack[c]{\rule{0pt}{10pt}29.53\\28.73\\1.03} & \shortstack[c]{\rule{0pt}{10pt}29.20\\28.62\\1.02} & \shortstack[c]{\rule{0pt}{10pt}29.09\\28.15\\1.03} & \shortstack[c]{\rule{0pt}{10pt}28.08\\28.15\\1.00} & \shortstack[c]{\rule{0pt}{10pt}6.33\\6.17\\1.03} & \shortstack[c]{\rule{0pt}{10pt}6.40\\5.82\\1.10} & \shortstack[c]{\rule{0pt}{10pt}5.55\\5.51\\1.01} & \shortstack[c]{\rule{0pt}{10pt}6.35\\5.95\\1.07} & \shortstack[c]{\rule{0pt}{10pt}5.72\\5.90\\0.97} & \shortstack[c]{\rule{0pt}{10pt}6.26\\5.90\\1.06} & 1.03\\
KappaCeti & \shortstack[c]{\rule{0pt}{10pt}29.61\\29.45\\1.01} & \shortstack[c]{\rule{0pt}{10pt}29.29\\29.26\\1.00} & \shortstack[c]{\rule{0pt}{10pt}29.18\\29.23\\1.00} & \shortstack[c]{\rule{0pt}{10pt}28.13\\29.23\\0.96} & \shortstack[c]{\rule{0pt}{10pt}6.38\\6.66\\0.96} & \shortstack[c]{\rule{0pt}{10pt}6.41\\6.33\\1.01} & \shortstack[c]{\rule{0pt}{10pt}5.56\\6.36\\0.87} & \shortstack[c]{\rule{0pt}{10pt}6.35\\6.20\\1.02} & \shortstack[c]{\rule{0pt}{10pt}5.50\\6.33\\0.87} & \shortstack[c]{\rule{0pt}{10pt}5.80\\6.33\\0.92} & 0.96\\
deltaPav & \shortstack[c]{\rule{0pt}{10pt}28.07\\28.22\\0.99} & \shortstack[c]{\rule{0pt}{10pt}27.64\\28.02\\0.99} & \shortstack[c]{\rule{0pt}{10pt}27.56\\27.87\\0.99} & \shortstack[c]{\rule{0pt}{10pt}27.24\\27.87\\0.98} & \shortstack[c]{\rule{0pt}{10pt}5.15\\5.35\\0.96} & \shortstack[c]{\rule{0pt}{10pt}5.13\\4.96\\1.04} & \shortstack[c]{\rule{0pt}{10pt}4.38\\5.07\\0.86} & \shortstack[c]{\rule{0pt}{10pt}5.26\\4.96\\1.06} & \shortstack[c]{\rule{0pt}{10pt}4.75\\4.98\\0.95} & -- & 0.98\\
55CncA & \shortstack[c]{\rule{0pt}{10pt}28.28\\28.23\\1.00} & \shortstack[c]{\rule{0pt}{10pt}27.87\\28.02\\0.99} & \shortstack[c]{\rule{0pt}{10pt}27.79\\27.96\\0.99} & \shortstack[c]{\rule{0pt}{10pt}27.37\\27.96\\0.98} & \shortstack[c]{\rule{0pt}{10pt}5.52\\5.36\\1.03} & \shortstack[c]{\rule{0pt}{10pt}5.27\\4.98\\1.06} & \shortstack[c]{\rule{0pt}{10pt}4.50\\5.12\\0.88} & \shortstack[c]{\rule{0pt}{10pt}5.38\\4.90\\1.10} & \shortstack[c]{\rule{0pt}{10pt}4.99\\4.99\\1.00} & \shortstack[c]{\rule{0pt}{10pt}4.90\\4.99\\0.98} & 1.00\\
GL892 & \shortstack[c]{\rule{0pt}{10pt}28.21\\28.52\\0.99} & \shortstack[c]{\rule{0pt}{10pt}27.80\\28.36\\0.98} & \shortstack[c]{\rule{0pt}{10pt}27.72\\28.27\\0.98} & \shortstack[c]{\rule{0pt}{10pt}27.33\\28.27\\0.97} & \shortstack[c]{\rule{0pt}{10pt}5.40\\5.75\\0.94} & \shortstack[c]{\rule{0pt}{10pt}5.30\\5.46\\0.97} & \shortstack[c]{\rule{0pt}{10pt}4.54\\5.41\\0.84} & \shortstack[c]{\rule{0pt}{10pt}5.41\\5.35\\1.01} & -- & \shortstack[c]{\rule{0pt}{10pt}4.95\\5.47\\0.91} & 0.95\\
GL183 & \shortstack[c]{\rule{0pt}{10pt}28.28\\27.50\\1.03} & \shortstack[c]{\rule{0pt}{10pt}27.87\\27.33\\1.02} & \shortstack[c]{\rule{0pt}{10pt}27.79\\27.08\\1.03} & \shortstack[c]{\rule{0pt}{10pt}27.37\\27.08\\1.01} & \shortstack[c]{\rule{0pt}{10pt}5.53\\5.04\\1.10} & \shortstack[c]{\rule{0pt}{10pt}5.45\\4.55\\1.20} & \shortstack[c]{\rule{0pt}{10pt}4.67\\4.56\\1.02} & \shortstack[c]{\rule{0pt}{10pt}5.53\\4.72\\1.17} & \shortstack[c]{\rule{0pt}{10pt}5.00\\4.65\\1.08} & \shortstack[c]{\rule{0pt}{10pt}4.80\\4.65\\1.03} & 1.07\\
GJ832 & \shortstack[c]{\rule{0pt}{10pt}27.78\\26.75\\1.04} & \shortstack[c]{\rule{0pt}{10pt}27.33\\26.62\\1.03} & \shortstack[c]{\rule{0pt}{10pt}27.26\\26.29\\1.04} & \shortstack[c]{\rule{0pt}{10pt}27.08\\26.29\\1.03} & \shortstack[c]{\rule{0pt}{10pt}5.53\\4.67\\1.18} & \shortstack[c]{\rule{0pt}{10pt}5.32\\4.30\\1.24} & \shortstack[c]{\rule{0pt}{10pt}4.56\\4.23\\1.08} & \shortstack[c]{\rule{0pt}{10pt}5.43\\4.38\\1.24} & \shortstack[c]{\rule{0pt}{10pt}4.74\\4.34\\1.09} & \shortstack[c]{\rule{0pt}{10pt}4.17\\4.34\\0.96} & 1.09\\
Lalande21185 (2001) & \shortstack[c]{\rule{0pt}{10pt}28.13\\27.40\\1.03} & \shortstack[c]{\rule{0pt}{10pt}27.71\\27.20\\1.02} & \shortstack[c]{\rule{0pt}{10pt}27.63\\26.00\\1.06} & \shortstack[c]{\rule{0pt}{10pt}27.28\\26.00\\1.05} & \shortstack[c]{\rule{0pt}{10pt}5.76\\5.60\\1.03} & \shortstack[c]{\rule{0pt}{10pt}5.79\\5.22\\1.11} & \shortstack[c]{\rule{0pt}{10pt}4.99\\5.02\\0.99} & \shortstack[c]{\rule{0pt}{10pt}5.83\\5.36\\1.09} & -- & -- & 1.05\\
Lalande21185 (2004) & \shortstack[c]{\rule{0pt}{10pt}29.16\\28.77\\1.01} & \shortstack[c]{\rule{0pt}{10pt}28.80\\28.76\\1.00} & \shortstack[c]{\rule{0pt}{10pt}28.70\\26.00\\1.10} & \shortstack[c]{\rule{0pt}{10pt}27.87\\26.00\\1.07} & \shortstack[c]{\rule{0pt}{10pt}6.24\\7.00\\0.89} & \shortstack[c]{\rule{0pt}{10pt}6.68\\6.80\\0.98} & \shortstack[c]{\rule{0pt}{10pt}5.81\\5.02\\1.16} & \shortstack[c]{\rule{0pt}{10pt}6.58\\6.92\\0.95} & -- & -- & 1.02\\ProxCen & \shortstack[c]{\rule{0pt}{10pt}28.01\\26.51\\1.06} & \shortstack[c]{\rule{0pt}{10pt}27.58\\26.38\\1.05} & \shortstack[c]{\rule{0pt}{10pt}27.50\\25.16\\1.09} & \shortstack[c]{\rule{0pt}{10pt}27.21\\25.16\\1.08} & \shortstack[c]{\rule{0pt}{10pt}6.24\\5.67\\1.10} & \shortstack[c]{\rule{0pt}{10pt}6.26\\5.28\\1.18} & \shortstack[c]{\rule{0pt}{10pt}5.42\\4.88\\1.11} & \shortstack[c]{\rule{0pt}{10pt}6.22\\5.43\\1.15} & \shortstack[c]{\rule{0pt}{10pt}5.44\\5.36\\1.02} & \shortstack[c]{\rule{0pt}{10pt}5.00\\5.36\\0.93} & 1.08\\
\enddata

\tablecomments{
Cells report:
top = log scaling prediction,
middle = log model value in same band,
bottom = scaling/model ratio.
Final column gives the mean ratio across all available relations.
}
\end{deluxetable*}


\input{output.bbl}
\end{document}

%% file: stellar_props_REVISED.tex
\begin{deluxetable*}{l l l l l l l l l l l}
\tablecaption{Stellar Properties}\label{tab:stellar_properties}
\tablehead{
\colhead{Name} &
\colhead{HD Name} &
\colhead{SpT} &
\colhead{$V_{\rm mag}$} &
\colhead{$T_{\rm eff}$ (K)} &
\colhead{$\log g$ (cgs)} &
\colhead{$M_\star$ ($M_\odot$)} &
\colhead{[Fe/H]} &
\colhead{$R_\star$ ($R_\odot$)} &
\colhead{Parallax (mas)} &
\colhead{TSS25 list?}
}
\startdata
iota Hor & HD 17051 & F9V & 5.4 &
$6298\pm50^{a}$ &
$4.56\pm0.05^{a}$ &
$1.18^{b}$ &
$0.20\pm0.05^{a}$ &
$1.13^{c}$ &
$57.61\pm0.04^{d}$ &
Y \\
$\chi^1$ Ori & HD 39587 & G0V & 4.4 &
$5906\pm92^{e}$ &
$4.39\pm0.14^{e}$ &
$1.06^{b}$ &
$-0.06\pm0.06^{e}$ &
0.956$^{s}$ &
$114.95\pm0.52^{d}$ &
N \\
GL311 ($\pi^1$ UMa) & HD 72905 & G1.5V & 5.64 &
$5962\pm100^{f}$ &
$4.47\pm0.10^{f}$ &
$1.01^{b}$ &
$-0.08\pm0.04^{f}$ &
$0.96^{c}$ &
$69.26\pm0.05^{d}$ &
Y \\
GL620-1A & HD 147513 & G1V & 5.376 &
$5818\pm35^{g}$ &
$4.48\pm0.02^{g}$ &
$1.03^{b}$ &
$0.09\pm0.01^{g}$ &
$0.97^{c}$ &
$77.57\pm0.07^{d}$ &
Y \\
$\kappa^1$ Cet & HD 20630 & G5V & 4.85 &
$5605\pm55^{g}$ &
$4.40\pm0.07^{g}$ &
$1.04^{b}$ &
$0.04\pm0.02^{g}$ &
$0.95^{c}$ &
$107.80\pm0.18^{d}$ &
Y \\
$\delta$ Pav & HD 190248 & G8IV & 3.56 &
$5454\pm92^{h}$ &
$4.03\pm0.17^{h}$ &
$0.94^{b}$ &
$0.20\pm0.07^{h}$ &
$1.19^{c}$ &
$163.95\pm0.12^{d}$ &
Y \\
55 Cnc A ($\rho^1$ Cnc) & HD 75732 & K0IV/V & 5.951 &
$5320\pm57^{a}$ &
$4.24\pm0.13^{a}$ &
$0.88^{b}$ &
$0.36\pm0.04^{a}$ &
$0.96^{c}$ &
$79.45\pm0.04^{d}$ &
Y \\
GL 892 & HD 219134 & K3V & 5.57 &
$4840\pm100^{j}$ &
$4.48\pm0.10^{j}$ &
$0.78^{b}$ &
$0.11\pm0.06^{j}$ &
$0.78^{c}$ &
$152.86\pm0.05^{d}$ &
Y \\
GL 183 & HD 32147 & K3V & 6.21 &
$4726\pm100^{i}$ &
$4.23\pm0.10^{i}$ &
$0.78^{b}$ &
$0.21\pm0.05^{i}$ &
$0.74^{c}$ &
$113.07\pm0.02^{d}$ &
Y \\
GJ 832 & HD 204961 & M1.5V & 8.672 &
$3657\pm185^{k}$ &
$4.7\pm0.1^{l}$ &
$0.47^{b}$ &
$-0.17\pm0.09^{m}$ &
$0.499\pm0.017^{n}$ &
$201.32\pm0.02^{d}$ &
N \\
Lalande 21185 & HD 95735 & M2V & 7.52 &
$3547\pm18^{o}$ &
4.84$^{s}$ &
$0.389\pm0.008^{o}$ &
$-0.362\pm0.12^{k}$ &
$0.392\pm0.004^{o}$ &
$391.5\pm0.03^{d}$ &
Y \\
Proxima Cen (GJ 551) & \nodata & M5.5V & 11.3 &
$2900\pm100^{p}$ &
$5.16\pm0.1^{q}$ &
$0.123^{b}$ &
$0.04\pm0.04^{p}$ &
$0.141\pm0.021^{p}$ &
$768.07\pm0.05^{d}$ &
N \\
\enddata
\tablecomments{
{\bf References:}
$^{a}$ \citealt{Perdelwitz2024};
$^{b}$ \citealt{Pecaut2013};
$^{c}$ \citealt{Hinkel2014};
$^{d}$ \citealt{Gaia2021};
$^{e}$ \citealt{Arentsen2019};
$^{f}$ \citealt{Garcia2021};
$^{g}$ \citealt{Boeche2016};
$^{h}$ \citealt{Bond2006};
$^{i}$ \citealt{Pal2023};
$^{j}$ \citealt{Zandt2023};
$^{k}$ \citealt{Bailey2009};
$^{l}$ \citealt{Wittenmyer2014};
$^{m}$ \citealt{Neves2014};
$^{n}$ \citealt{Houdebine2010};
$^{o}$ \citealt{Pineda2021};
$^{p}$ \citealt{Faria2022};
$^{q}$ \citealt{Stassun2018};
$^{r}$ \citealt{Soubiran2024};
$^{s}$ This work.
}

\end{deluxetable*}

%% file: table_observations.tex
\begin{deluxetable*}{lllccl}
\tablecaption{Ultraviolet Observations Used in This Work \label{tab:obs}}
\tablehead{
\colhead{Star (HD)} & \colhead{Band} & \colhead{Instrument/Mode/$\lambda_c$} & \colhead{Date} & \colhead{PID}
}
\startdata
$\iota$ Hor (HD 17051) & FUV & STIS/E140M/1425\,\AA & 2019-08-01 & 15512 \\
                       & NUV & STIS/E230H/2713\,\AA & 2019-08-01 & 15512 \\
\hline
$\chi^1$ Ori (HD 39587)$^{a}$ & FUV & STIS/E140M/1425\,\AA & 2000-10-03 & 8280 \\
                        & NUV & STIS/E230M/2124\,\AA & 2000-10-03 & 8280 \\
\hline
GL 311 (HD 72905) & FUV & STIS/E140M/1425\,\AA & 2012-09-25 & 12596 \\
                  & NUV & STIS/E230H/2713\,\AA & 2010-04-24 & 11568 \\
\hline
GL 620-1A (HD 147513) & FUV & STIS/E140M/1425\,\AA & 2018-08-30, 2019-07-14 & 15299 \\
                      & NUV & STIS/E230H/2713\,\AA & 2018-08-30 & 15299 \\
\hline
$\kappa^1$ Cet (HD 20630)$^{a}$ & FUV & STIS/E140M/1425\,\AA & 2000-09-19 & 8280 \\
                          & NUV & STIS/E230M/2124\,\AA & 2000-09-19 & 8280 \\
                          & NUV & STIS/E230H/2862\,\AA & 2000-09-19 & 8280 \\
                          & NUV & STIS/E230H/2663\,\AA & 2000-09-19 & 8280 \\
\hline
$\delta$ Pav (HD 190248) & FUV & STIS/E140H/1271\,\AA & 2015-09-19 & 13658 \\
                         & NUV & STIS/E230H/2713\,\AA & 2015-09-19 & 13658 \\
\hline
55 Cnc A (HD 75732) & FUV & STIS/G140M/1222\,\AA & 2012-03-07, 2012-04-05 & 12681 \\
                    & FUV & COS/G130M/1291\,\AA & 2016-04-04 & 14094 \\
                    & NUV & STIS/E230M/2707\,\AA & 2016-04-17, 2016-04-24 & 14094 \\
\hline
GL 892 (HD 219134) & FUV & STIS/E140H/1271\,\AA & 2018-02-16, 2018-07-02 & 15430 \\
                   & NUV & STIS/G140M/1222\,\AA & 2016-10-15 & 14461 \\
                   & NUV & STIS/E230H/2713\,\AA & 2020-11-18 & 16225 \\
\hline
GL 183 (HD 32147) & FUV & STIS/E140M/1425\,\AA & 2016-09-16 & 14084 \\
                  & NUV & STIS/E230H/2713\,\AA & 2016-09-16 & 14084 \\
\hline
GJ 832 (HD 204961)$^{b}$ & FUV & STIS/G140M/1222\,\AA & 2014-10-10 & 13650 \\
                   & FUV & COS/G130M/1291, 1318\,\AA & 2014-10-11 & 13650 \\
                   & NUV & COS/G160M/1577, 1611\,\AA & 2014-10-10 & 13650 \\
                   & NUV & STIS/G230L/2376\,\AA & 2014-10-10 & 13650 \\
                   & NUV & COS/G230L/2950\,\AA & 2014-10-10, 2014-10-11 & 13650 \\
\hline
Lalande 21185 (HD 95735)$^{c}$ & FUV & STIS/E140M/1425\,\AA & 2019-05-15, 2020-01-14 & 15190 \\
                         & NUV & STIS/G230LB/2375\,\AA & 2002-02-08 & 9088 \\
\hline
Proxima Cen$^{b}$ & FUV & STIS/E140M/1425\,\AA & 2000-05-08, 2000-05-09 & 8040 \\
            & NUV & STIS/E230H/2713\,\AA & 2022-08-16 & 16225 \\
\enddata
\tablenotetext{a}{StarCAT spectral energy distribution \citep{Ayres2010}.}
\tablenotetext{b}{MUSCLES spectral energy distribution, v22 var-res-sed.fits data products \citep{France2016,Youngblood16,Loyd2016}. We use selected wavelength regions from these products: 1205--1740 and 2250--5500~\AA\ for GJ~832, and select FUV emission lines (including the Ly$\alpha$ reconstruction) and 2250--3050~\AA\ for Proxima Cen.}
\tablenotetext{c}{STIS Next Generation Stellar Library (NGSL, version 1) G230LB spectrum \citep{Gregg2004}.}
\end{deluxetable*}

%% file: obs_epoch_table.tex
\begin{deluxetable*}{lcccc}
\tablecaption{Observational Epoch Summary\label{tab:epoch_summary}}
\tablehead{
\colhead{Star} &
\colhead{X-ray Epoch(s)} &
\colhead{FUV Epoch(s)} &
\colhead{NUV Epoch(s)} &
\colhead{$\Delta t$ (yr)}
}
\startdata
$\iota$ Hor 
& 2011-05-16 -- 2018-02-03 (32) 
& 2019-08-01 
& 2019-08-01 
& 1--8 \\
$\chi^1$ Ori 
& 2001-04-07
& 2000-10-03 
& 2000-10-03 
& 0.5 \\
GL 311 
& 2000-11-04 
& 2012-09-25 
& 2010-04-24 
& 10--12 \\
GL 620-1A 
& 2018-09-08 
& 2018-08-30, 2019-07-14 
& 2018-08-30 
& $<1$ \\
$\kappa^1$ Cet 
& 2002-02-09 (1), 2018-07-30 -- 2021-07-22 (4)
& 2000-09-19 
& 2000-09-19 
& 2--21 \\ 
& 2019-10-22, 2019-11-18 (Chandra)
& 
& 
& \\
$\delta$ Pav 
& 2016-10-05 
& 2015-09-19 
& 2015-09-19 
& 1 \\
55 Cnc A 
& 2009-04-11 
& 2012-03-07, 2012-04-05; 2016-04-04 
& 2016-04-17, 2016-04-24 
& 3--7 \\
GL 892 
& 2016-06-13 
& 2018-02-16, 2018-07-02 
& 2016-10-15; 2020-11-18 
& 0--4 \\
GL 183 
& 2017-03-16 
& 2016-09-16 
& 2016-09-16 
& 0.5 \\
GJ 832 
& 2014-10-11, 2021-04-14 
& 2014-10-10, 2014-10-11 
& 2014-10-10, 2014-10-11 
& 0--7 \\
Lalande 21185
& 2001-05-15, 2004-05-12 
& 2019-05-15, 2020-01-14 
& 2002-02-08 
& 1--18 \\
Proxima Cen 
& 2001-08-12 -- 2018-03-11 (9) 
& 2000-05-08, 2000-05-09 
& 2022-08-16
& $<$1--23 \\
& 2000-05-07, 2000-05-08 (Chandra)
&
&
&
\enddata
\tablenotetext{}{X-ray observations are primarily from XMM-Newton, with additional Chandra observations for $\kappa^1$ Cet and Proxima Centauri listed in secondary rows and noted with parentheses. For stars with multiple X-ray observations we list the date range and total number of observations.}
\tablenotetext{}{ $\Delta t$ indicates the approximate temporal separation between the UV and X-ray observations. For stars with multiple epochs we list the minimum and maximum separations. Program IDs for X-ray observations are provided in \citet{Binder2024}.}
\end{deluxetable*}